\documentclass[11pt,aps,prd,english,nofrontpage,floatfix,nofootinbib,superscriptaddress,aps,prd,showkeys]{revtex4}
\usepackage[utf8]{inputenc}
\usepackage{float}
\usepackage{array}
\usepackage{lipsum}
\usepackage{graphicx}
\usepackage{amsmath,amsthm,amsfonts,amssymb}
\usepackage{tikz}
\usepackage{multirow}

\newcommand{\be}{\begin{equation}}
\newcommand{\ee}{\end{equation}}
\newcommand{\Be}{\begin{eqnarray}}
\newcommand{\Ee}{\end{eqnarray}}

\newcommand{\mincir}{\raise
-3.truept\hbox{\rlap{\hbox{$\sim$}}\raise4.truept\hbox{$<$}\ }}
\newcommand{\magcir}{\raise
-3.truept\hbox{\rlap{\hbox{$\sim$}}\raise4.truept\hbox{$>$}\ }}

\newcolumntype{Y}{>{\centering\arraybackslash}X}
\providecommand{\U}[1]

 \usetikzlibrary{quotes,angles}
 \usetikzlibrary{arrows} 
\usetikzlibrary{decorations.markings}
\usepackage{graphicx}
\usepackage[english]{babel}
\usepackage{color}
\usepackage{subfig}
\usepackage{caption}
\usepackage{tensor}
\usepackage{esint}
\usepackage[dvips]{epsfig}
\usepackage[dvips]{graphicx}
\usepackage{float}
\usepackage{units}
\usepackage{textcomp}
\usepackage{mathrsfs}
\usepackage{amsmath}
\usepackage[makeroom]{cancel}
\usepackage{amssymb}
\usepackage{amsbsy}
\usepackage{amsfonts}
\usepackage{amssymb,mathrsfs,xcolor}
\usepackage{esint}
\usepackage{array}
\usepackage{graphicx}

\usepackage{hyperref}
\hypersetup{
    colorlinks,
    citecolor=blue,
    filecolor=green,
    linkcolor=purple,
    urlcolor=red,
}

\usepackage{slashed}

\newcommand{\ie}{\begin{equation}}
\newcommand{\fe}{\end{equation}}
\newcommand{\se}{\begin{eqnarray}}
\newcommand{\ff}{\end{eqnarray}}

\usepackage{hyperref}
\hypersetup{colorlinks,breaklinks,
			citecolor=[rgb]{0,0.0,1.0},
            urlcolor=[rgb]{0.0,0.0,1.0},
            linkcolor=[rgb]{0,0.5,0.9}}

\begin{document}

\title{Properties of an axisymmetric Lorentzian non--commutative black hole}

\author{A. A. Ara\'{u}jo Filho}
\email{dilto@fisica.ufc.br}
\affiliation{Departamento de Física, Universidade Federal da Paraíba, Caixa Postal 5008, 58051-970, João Pessoa, Paraíba,  Brazil}

\author{J. R. Nascimento}
\email{jroberto@fisica.ufpb.br}
\affiliation{Departamento de Física, Universidade Federal da Paraíba, Caixa Postal 5008, 58051-970, João Pessoa, Paraíba,  Brazil}

\author{A. Yu. Petrov}
\email{petrov@fisica.ufpb.br}
\affiliation{Departamento de Física, Universidade Federal da Paraíba, Caixa Postal 5008, 58051-970, João Pessoa, Paraíba,  Brazil}

\author{P. J. Porfírio}
\email{pporfirio@fisica.ufpb.br}
\affiliation{Departamento de Física, Universidade Federal da Paraíba, Caixa Postal 5008, 58051-970, João Pessoa, Paraíba,  Brazil} 

\author{Ali \"Ovg\"un}
\email{ali.ovgun@emu.edu.tr}
\affiliation{Physics Department, Eastern Mediterranean
University, Famagusta, 99628 North Cyprus, via Mersin 10, Turkiye}


\date{\today}

\begin{abstract}

In this work, we start by examining a spherically symmetric black hole within the framework of non--commutative geometry and apply a modified Newman--Janis method to obtain a new rotating solution. We then investigate its consequences, focusing on the horizon structure, ergospheres, and the black hole's angular velocity. Following this, a detailed thermodynamic analysis is performed, covering surface gravity, \textit{Hawking} temperature, entropy, and heat capacity. We also study geodesic motion, with particular emphasis on null geodesics and their associated radial accelerations. Additionally, the photon sphere and the resulting black hole shadows are explored. Finally, we compute the \textit{quasinormal} modes for scalar perturbations using the 6th--order WKB approximation.

\end{abstract}

\keywords{Black holes; Noncommutative geometry;  Thermodynamics; Shadows; Quasinormal modes. }

\maketitle

\section{Introduction}
\label{introduction}

Non--commutative effective theories are well--known to arise from the low--energy limits of string/M-theory \cite{t15}. However, the concept of non--commutativity has a long history, particularly in quantum mechanics, where it is embodied in the uncertainty principle. In field theory, it can be introduced by allowing spacetime coordinates to no longer commute, leading to specific non--commutative relations. Various methods have been developed to formalize this concept. A common approach is the Moyal product, which is useful in the perturbative treatment of non--commutative field theories (see \cite{Girotti:2003at} for a review and \cite{Pernici:2000va, Zanon:2000nq, Ferrari:2003vs, Ferrari:2004ex} for examples). Another approach, particularly suitable for tree--level analysis, is the Seiberg--Witten (SW) map \cite{t15}, which we employ in this paper for studying gravity in a non--commutative setting. While a consistent way to apply the Moyal product in curved spacetime is yet to be established, the SW map offers a straightforward framework for our purposes here.

The SW method has been extensively used to study black hole thermodynamics and evaporation, resulting in modified expressions for temperature and entropy in the presence of non--commutative geometry \cite{chamseddine2001deforming, myung2007thermodynamics, touati2024quantum, heidari2024quantum, nozari2007thermodynamics, banerjee2008noncommutative, sharif2011thermodynamics}. By replacing point--like mass distributions with smooth profiles, such as Gaussian or Lorentzian functions, it avoids singularities and has contributed to investigate black hole radiation, shadows, and gravitational lensing \cite{Vagnozzi:2022moj,nicolini2006noncommutative, ghosh2018noncommutative, sharif2011thermodynamics, ovgun2020shadow, wei2015shadow, lekbich2024optical, saleem2023observable, ding2011strong, ding2011probing}. Non--commutative effects are also considered as small perturbative corrections in gravitational models, further expanding their application to astrophysical phenomena \cite{newcommutativity, herceg2024noncommutative, herceg2024metric}. Estimates for the noncommutative parameter have been obtained within the framework of particle creation \cite{touati2024quantum}, resulting in \(\Theta \approx 1.23585 \times 10^{-35} \, \text{m} \sim l_{\text{Planck}}\). This result supports the notion that the non--commutative parameter lies at the Planck scale, consistent with outcomes from gravitational wave observations \cite{rev45} and thermodynamic investigations \cite{rev37}. Additionally, various studies \cite{rev37,rev46,rev47,rev48,rev49} have constrained \(\Theta\) using black hole radiation, often approximating \(\sqrt{\Theta} \sim 10^{-1} l_{\text{Planck}}\). In Refs. \cite{rev44,rev50}, there was estimated a lower bound for \(\Theta\) in the range of \(10^{-31} \, \text{m}\) to \(10^{-34} \, \text{m}\). However, thermodynamic approaches \cite{rev37} provided a more accurate estimation of \(\Theta \sim 10^{-35} \, \text{m}\).

On the other hand, gravitational waves have become fundamental for studying strong--field gravity and compact objects. These waves carry information about dynamic events like black hole mergers or stellar oscillations \cite{unno1979nonradial, kjeldsen1994amplitudes, dziembowski1992effects, pretorius2005evolution, hurley2002evolution, yakut2005evolution, heuvel2011compact}. Particularly important are quasinormal modes--oscillations that reveal the stability and internal structure of black holes \cite{kokkotas1999quasi,roy2020revisiting,oliveira2019quasinormal, berti2009quasinormal, horowitz2000quasinormal, heidari2024impact,Hamil:2024ppj,nollert1999quasinormal,ferrari1984new,santos2016quasinormal,ovgun2018quasinormal,jusufi2024charged, rincon2020greybody,araujo2024dark,london2014modeling,maggiore2008physical, flachi2013quasinormal,blazquez2018scalar,konoplya2011quasinormal}. The study of these modes has been extended to alternative gravity theories and cases involving Lorentz symmetry breaking, providing a broader picture of gravitational interactions \cite{Pedrotti:2024znu,gogoi2023quasinormal,mmm2,lee2020quasi, jawad2020quasinormal,maluf2013matter, maluf2014einstein,jcap4,JCAP1,Daghigh:2008jz,JCAP2,JCAP3, jcap5,hassanabadi2023gravitational,Heidari:2023bww,mmm1,Fernando:2016ftj,araujo2024exploring,liu2022quasinormal,yang2023probing,Gogoi:2023fow,lambiase2023investigating,Fernando:2012yw,kim2018quasi,Daghigh:2020jyk,Daghigh:2005ph,Daghigh:2020mog,Daghigh:2022uws,Gogoi:2023kjt,Daghigh:2020fmw,Yang:2022ifo,Lambiase:2023zeo}.

A widely recognized and frequently employed method for generating rotating black hole solutions from spherically symmetric metrics is the Newman--Janis technique, originally introduced to incorporate angular momentum into otherwise static spacetimes \cite{n57,n57i}. This method utilizes a complex coordinate transformation, which has played a key role in deriving prominent solutions, such as the Kerr metric, from simpler static models like the Schwarzschild metric. In response to certain limitations of the original procedure, recent studies have developed several modifications to expand its applicability to a broader range of scenarios \cite{afrim}.

In addition, a non--complexification variant of the Newman--Janis approach, known as the modified Newman--Janis algorithm \cite{n58,n59}, offers an alternative method that bypasses the traditional complexification process, which has been a subject of ongoing debate. Instead, this approach relies on a more geometrically intuitive transformation. This modification has proven to be highly effective in deriving rotating metrics for systems involving imperfect fluids, expanding its applicability beyond the perfect fluid cases commonly addressed in the original framework. Through the use of this modified technique, it becomes possible to generate rotating solutions from static, spherically symmetric metrics across a wider array of physical scenarios, significantly increasing its versatility \cite{afrim,n63,n60,n65,n62,n61,n64,pantig2023testing}.

This study begins by examining a spherically symmetric black hole within the framework of non--commutative geometry, employing the corrected Newman--Janis algorithm to derive a new rotating solution in this scenario. The analysis then shifts to the key features of the solution, including the horizons, ergospheres, and angular velocity of the black hole. A comprehensive thermodynamic investigation follows, where we evaluate the surface gravity, \textit{Hawking} temperature, entropy, and heat capacity. Furthermore, we investigate the geodesics, paying particular attention to null geodesics and their associated radial accelerations. The photon sphere and the black hole’s shadow are explored. Finally, we compute the \textit{quasinormal} modes via the $6$--th order WKB approximation.

This work is organized as follows:  in Sec. \ref{cnj}, we discuss the modified Newman--Janis technique used to derive a new rotating solution in the framework of non--commutative geometry. In Sec. \ref{thermoanalysis}, we outline the key features of the solution and analyze its thermodynamic properties. In Sec. \ref{geoss}, we derive the geodesic equations. In Sec. \ref{shaaa}, we calculate the photon sphere and the resulting black hole shadows. In Sec. \ref{quasi}, we compute the quasinormal modes for scalar perturbations using the WKB approximation. Finally, in Sec. \ref{cccon}, we present our concluding remarks and final considerations.


\section{A corrected Newman--Janis technique \label{cnj}}

Exploring the implications of spacetime often involves incorporating the principles of non--commutativity within the framework of general relativity \cite{k10,k101,k102,k103,k6,k7,k8,k9,campos2022quasinormal,anacleto2021quasinormal,anacleto2023absorption}. Several formulations of non--commutative field theory, particularly those based on the Moyal product, have been developed to address this integration \cite{k11}. In this section, we begin by showing the essential characteristics of the static black hole solution under consideration. The mass density distribution is described by the following expression  $\rho_{\Theta}(r) = \frac{M \sqrt{\Theta}}{\pi^{3/2} (r^{2} + \pi \Theta)^{2}}$ \cite{Anacleto:2019tdj,campos2022quasinormal,nicolini2006noncommutative,nozari2008hawking}, where $M$ represents the total mass and $\Theta$ is the non--commutative parameter with the dimensions of $ \mathrm{L}^{2}$. The non--commutativity is defined as:
$[x^{\mu}, x^{\nu}] = i \Theta^{\mu \nu}$.
Additionally, we define the mass function $ \mathcal{M}_{\Theta} $ as:
$\mathcal{M}_{\Theta} = \int^{r}_{0} 4\pi r^{2} \rho_{\Theta}(r) \mathrm{d}r = M - \frac{4M \sqrt{\Theta}}{\sqrt{\pi}r}$.
In this non--commutative framework, the Schwarzschild--like black hole metric takes the form:
\ie
\label{staticmetricc}
\mathrm{d}s^{2} = -f_{\Theta}(r) \mathrm{d}t^{2} + f_{\Theta}(r)^{-1} \mathrm{d}r^{2} + r^{2} \mathrm{d}\theta^{2} + r^{2} \sin^{2} \theta \, \mathrm{d}\varphi^{2},
\fe
where the metric function $f_{\Theta}(r) $ is given by
$f_{\Theta}(r) = 1 - \frac{2M}{r} + \frac{8M\sqrt{\Theta}}{\sqrt{\pi}r^{2}}$ \cite{Anacleto:2019tdj,campos2022quasinormal}. It is worth noting that recent studies have thoroughly investigated the gravitational properties of this metric, including analyses of quasinormal modes, shadow behaviors, geodesics, gravitational lensing, thermodynamics, emission rates, and the black hole evaporation process \cite{nascimento2024effects,campos2022quasinormal}.

A well--established method for deriving rotating black hole solutions from spherically symmetric metrics is the Newman--Janis technique \cite{n57,n57i}. In this study, we apply a modified version of this technique, known as the non-complexification Newman--Janis algorithm \cite{n58,n59}. This adaptation avoids the complexification step and has proven particularly effective in obtaining rotating metrics for systems involving imperfect fluids, derived from static, spherically symmetric metrics \cite{n60,n65,n62,n61,n63,n64}.

The process begins by rewriting the given metric using advanced null coordinates, particularly the Eddington--Finkelstein coordinates \((u, r, \theta, \phi)\). This is accomplished by implementing the transformation:
$
\mathrm{d}u = \mathrm{d}t - \frac{\mathrm{d}r}{f_{\Theta}(r)}$,
which allows the static metric in Eq. (\ref{staticmetricc}) to be reformulated as follows
\ie
\mathrm{d} s^{2} = - f_{\Theta}(r) \mathrm{d}u^{2} - 2\mathrm{d}u \mathrm{d}r + r^{2}\left(  \mathrm{d}\theta^{2} + \sin^{2}\theta \mathrm{d}\phi^{2}   \right).
\fe
A set of null tetrad vectors is introduced, denoted as \( Z^\alpha_\mu = (l_\mu, n_\mu, m_\mu, \bar{m}_\mu) \), with the inverse metric tensor \( g^{\mu\nu} \) represented by the following expression:
$
g^{\mu\nu} = -l^\mu n^\nu - l^\nu n^\mu + m^\mu \bar{m}^\nu + m^\nu \bar{m}^\mu$.
The individual components of the tetrad are defined as follows:
$ l^{\mu} = \delta^{\mu}_{r}$, $ n^\mu = \delta^u_\mu - \frac{1}{2} f_{\Theta}(r) \delta^\mu_r$,  $m^\mu = \frac{1}{\sqrt{2}r} \left(\delta_\theta^\mu + \frac{i} {\sin\theta} \, \delta_\phi^\mu\right)$.
In this setting, \(\bar{m}_\mu\) corresponds to the complex conjugate of \(m_\mu\). The null tetrad vectors create an orthonormal basis and adhere to the following relationships: $
l^\mu l_\mu = n^\mu n_\mu = m^\mu m_\mu = l^\mu m_\mu = n^\mu m_\mu = 0$,
and
$ l^\mu n_\mu = -m^\mu \bar{m}_\mu = -1$.

Next, a complex coordinate transformation is applied, modifying the vectors \(\delta^\nu_\mu\) according to the following transformation \cite{afrim}:
\ie
\delta^r_\mu \rightarrow \delta^r_\mu, \quad \delta^u_\mu \rightarrow \delta^u_\mu, \quad \delta^\theta_\mu \rightarrow \delta^r_\mu + i a \sin \theta (\delta^u_\mu - \delta^r_\mu), \quad \delta^\phi_\mu \rightarrow \delta^\phi_\mu.
\fe
In this framework, \(a\) refers to the spin parameter of the black hole. The ambiguity surrounding the complexification of the radial coordinate is resolved through the modified Newman--Janis algorithm introduced by Azreg--Ainou \cite{azreg2014generating,afrim}. Instead of applying direct complexification to \(r\), this method alters the function \(f(r)\) into \(F(r, a, \theta)\) and replaces \(r^2\) with \(H(r, a, \theta)\). By adhering to the steps provided in \cite{azreg2014generating,afrim}, the null tetrads after transformation are derived as follows:
\ie
l'_\mu = \delta_r^\mu, \quad n'^\mu = \delta_u^\mu - F(r, a, \theta) \delta_r^\mu, \quad m'^\mu = \frac{1}{\sqrt{2H(r, a, \theta)}} \left[i a \sin\theta (\delta_u^\mu - \delta_r^\mu) + \delta_\theta^\mu + \sin\theta \, \delta_\phi^\mu \right],
\fe
Thus, with all the preliminaries established, we obtain the following result:
\ie
g^{\mu\nu} = -l'^\mu n'^\nu - l'^\nu n'^\mu + m'^\mu \bar{m}'^\nu + m'^\nu \bar{m}'^\mu.
\fe

Furthermore, using this approach, a rotating black hole configuration can be generated in the Eddington--Finkelstein coordinates, as shown below \cite{azreg2014generating}
\ie
\begin{split}
\mathrm{d}s^2 = &  -F(r, a, \theta) \mathrm{d}u^2 - 2 \mathrm{d}u \mathrm{d}r + 2 a \sin^2\theta \left(F(r, a, \theta) - 1\right) \mathrm{d}u \mathrm{d}\phi + 2 a \sin^2\theta \mathrm{d}r \mathrm{d}\phi \\ + & H(r, a, \theta) \mathrm{d}\theta^2 
+ \sin^2\theta \left[H(r, a, \theta) + a^2 \sin^2\theta \left(2 - F(r, a, \theta)\right)\right] \mathrm{d}\phi^2,
\end{split}
\fe
where \(F(r, a, \theta)\) represents a function derived from \(f_{\Theta}(r)\). Moreover, this method is highly adaptable and can be applied to any spherically symmetric solution to produce rotating spacetime configurations \cite{kumar2020rotating,brahma2021testing,islam2023investigating,afrin2022testing}. The crucial final step is to transform of the metric into Boyer--Lindquist coordinates, which is achieved through a global coordinate transformation \cite{azreg2014generating}
$\mathrm{d}u = \mathrm{d}t' + \lambda(r) \mathrm{d}r $, and $ \mathrm{d}\phi = \mathrm{d}\phi' + \chi(r) \mathrm{d}r$.
The functions \(\lambda(r)\) and \(\chi(r)\) depend entirely on the radial coordinate \(r\), and their explicit forms are determined as follows \cite{azreg2014generating}:
$\lambda(r) = -\frac{r^2 + a^2}{f_{\Theta}(r)(r^2 + a^2)}$, and $  \chi(r) = -\frac{a}{f_{\Theta}(r)(r^2 + a^2)}$.

\begin{figure}
    \centering
     \includegraphics[scale=0.5]{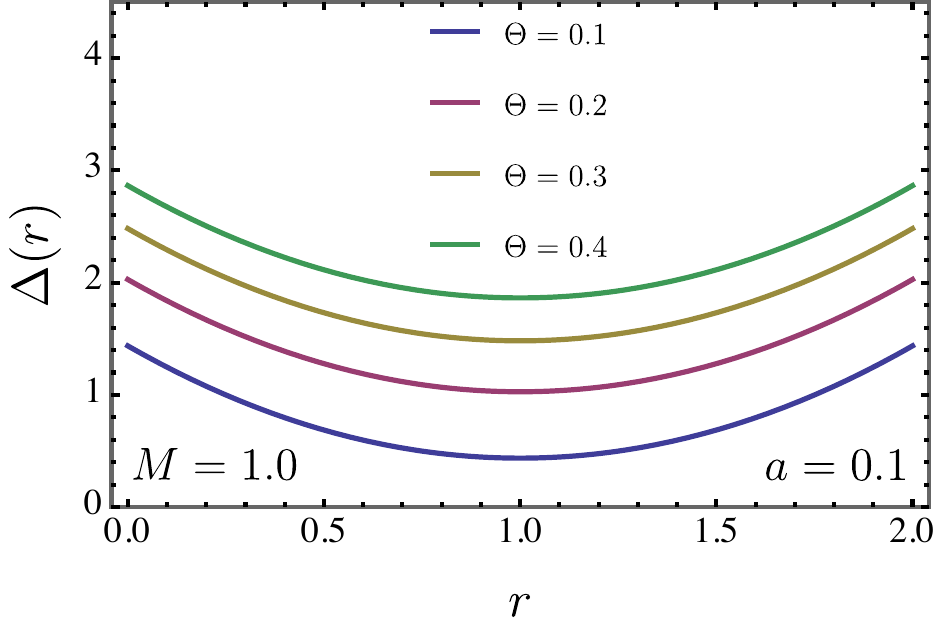}
     \includegraphics[scale=0.514]{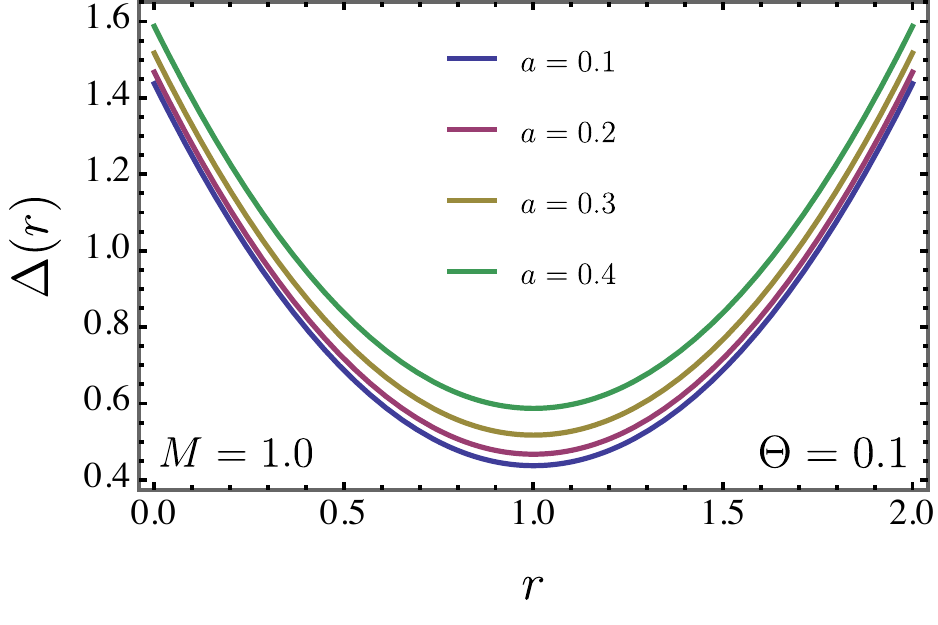}
    \caption{Parameter $\Delta(r)$ as a functions of $r$ for different configuration of $\Theta$ and $a$}
    \label{deltas}
\end{figure}

In order to remove the \(\mathrm{d}t \mathrm{d}r\) cross--term from the metric, the following condition is introduced: \(F(r,a,\theta) = \frac{f_{\Theta}(r)r^2 + a^2\cos^2\theta}{H(r,a,\theta)}\) \cite{afrim}. Additionally, for the Einstein tensor component \(G_{r\theta}\) to vanish, the condition \(H(r,a,\theta) = r^2 + a^2\cos^2\theta\) must be satisfied \cite{azreg2014generating}. After performing a series of algebraic manipulations, the resulting metric for a rotating black hole in Boyer–Lindquist coordinates is obtained as follows \cite{afrim}
\ie
\begin{split}
\label{rotatingmetric}
\mathrm{d} s^{2} = & \left[  \frac{\Delta(r) - a^{2} \sin^{2}\theta       }{\Sigma}   \right] \mathrm{d}t^{2} + \frac{\Sigma}{\Delta(r)} \mathrm{d}r^{2} + \Sigma \,\mathrm{d}\theta^{2} - 2 a \sin^{2} \theta \left[ 1 - \frac{\Delta(r) - a^{2}\sin^{2}\theta}{\Sigma}      \right] \mathrm{d}t \mathrm{d}\phi \\
& + \frac{\sin^{2}\theta}{\Sigma} \left[ (r^{2} + a^{2})^{2}  - \Delta(r) a^{2} \sin^{2} \theta   \right] \mathrm{d}\phi^{2},
\end{split}
\fe
where $\Delta(r) = a^{2} + r^{2} f_{\Theta}(r)$  and $\Sigma = r^{2} + a^{2} \cos^{2}\theta$. Given these preliminaries, we have successfully derived the rotating counterpart of the initial metric (\ref{staticmetricc}). Since the main characteristics are encapsulated by $\Delta(r)$, we now examine this function in Fig. \ref{deltas}. In this figure, we plot \(\Delta(r)\) as a function of \(r\), considering various values of \(\Theta\) and \(a\).
In the following sections, we shall examine the key characteristics of this newly obtained metric.


\section{General features and thermodynamics \label{thermoanalysis}}

\subsection{Horizons and ergosheres}

\begin{figure}
    \centering
     \includegraphics[scale=0.38]{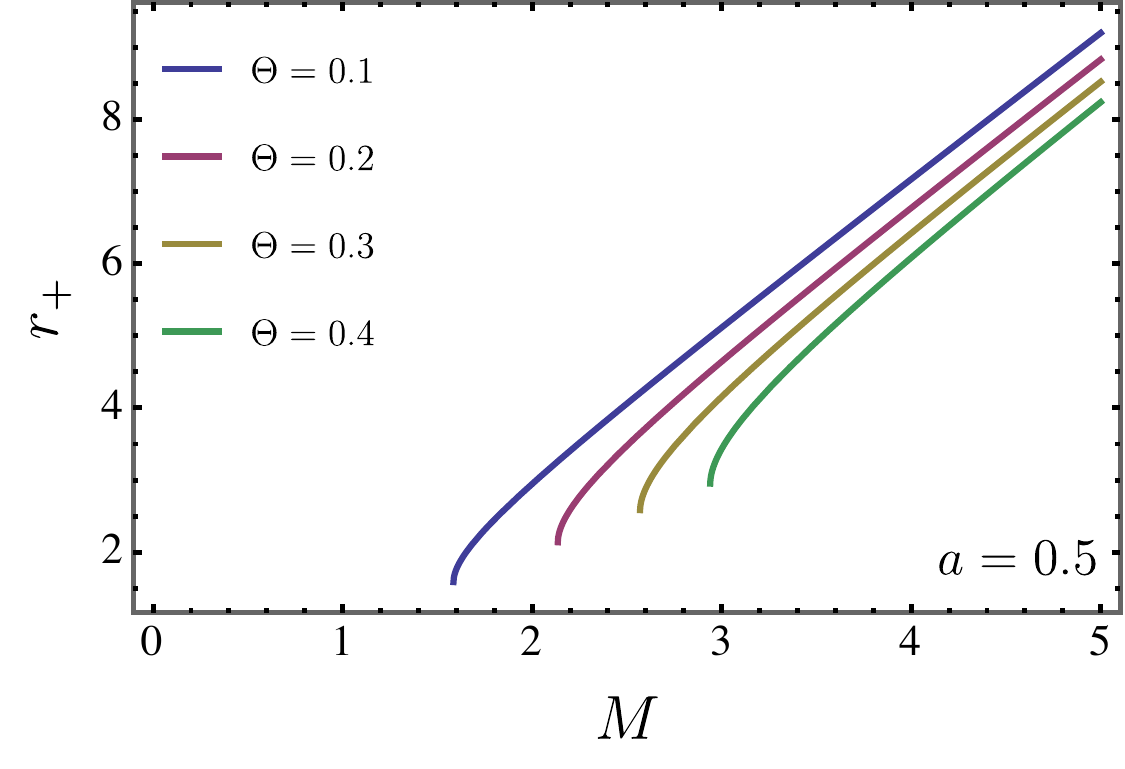}
     \includegraphics[scale=0.4]{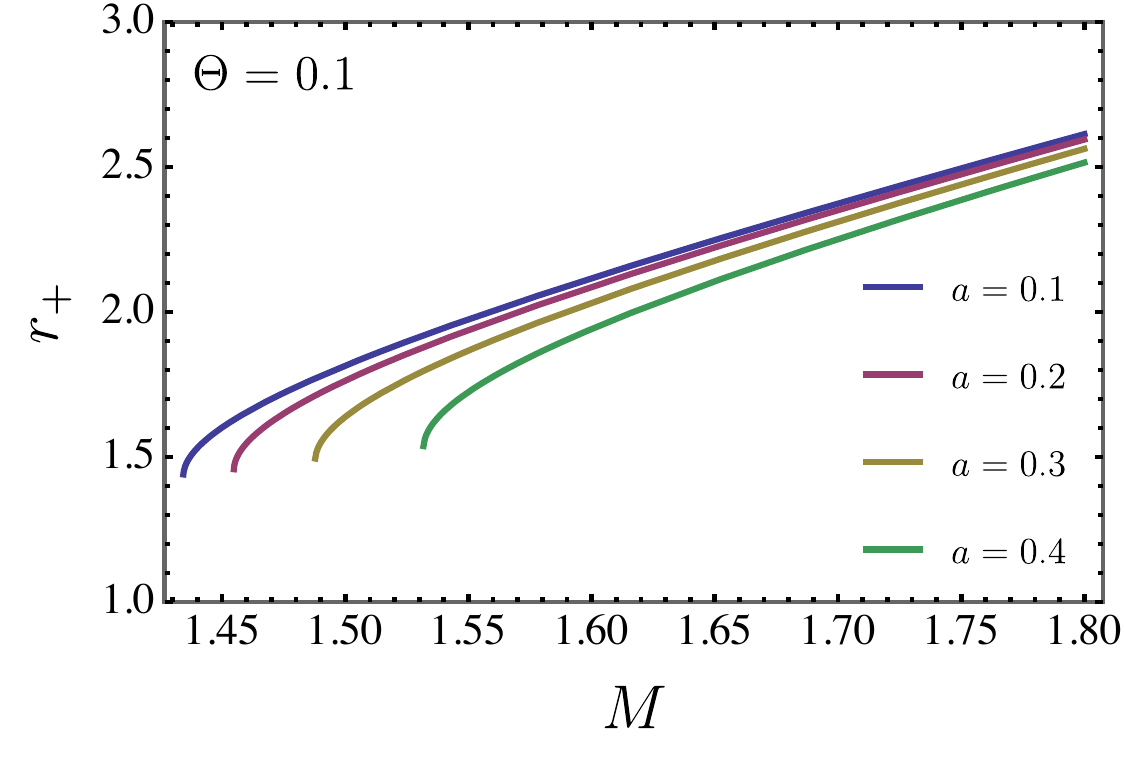}
     \includegraphics[scale=0.5]{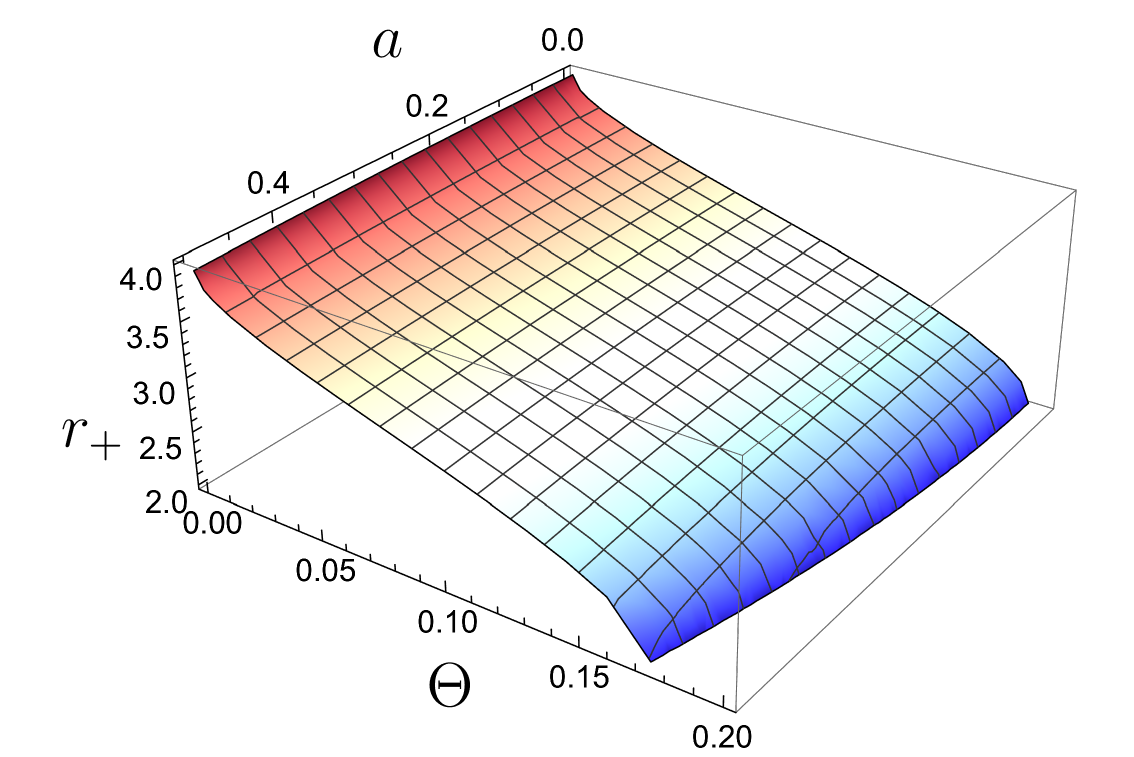}
    \caption{In the top left panel, the variation of \( r_{+} \) with \( M \) is shown for different values of \( \Theta \), with \( a = 0.5 \) held constant. The top right panel illustrates how \( r_{+} \) changes with \( M \) for varying values of \( a \), while keeping \( \Theta = 0.1 \) fixed. The bottom panel provides a three--dimensional plot of \( r_{+} \) as a function of \( a \) and \( \Theta \), with \( M = 2 \) held constant.}
    \label{eventhor}
\end{figure}

\begin{figure}
    \centering
     \includegraphics[scale=0.65]{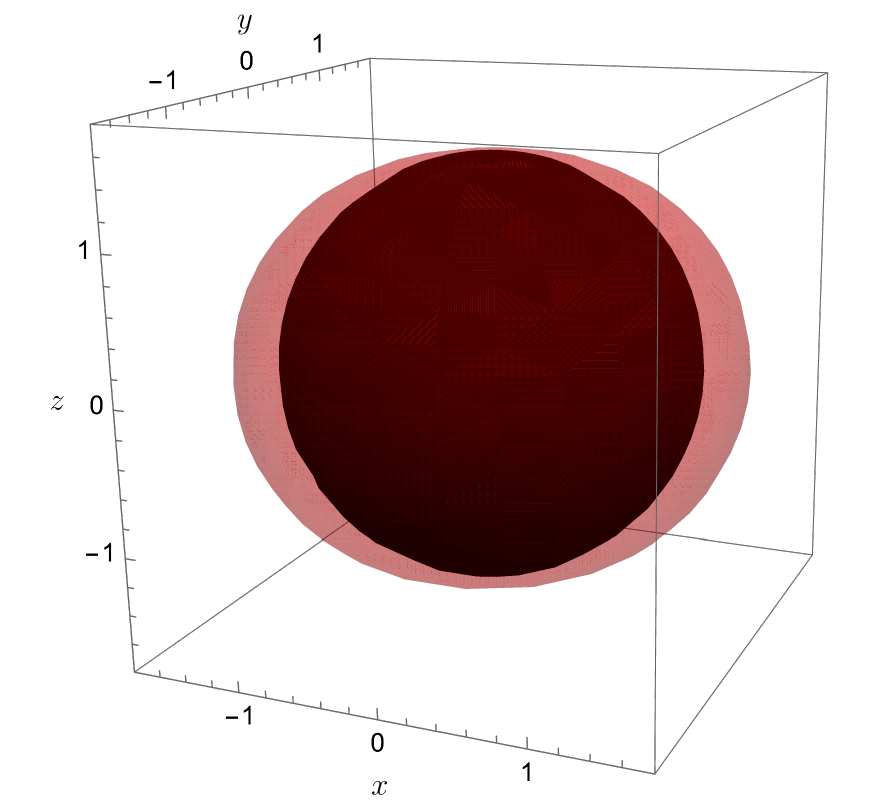}
    \caption{A 3--dimensional parametric visualization of the event horizon and ergosphere for the non--commutative black hole under consideration.}
    \label{ergosasdasd}
\end{figure}

In this section, in addition to exploring the thermal properties of the metric (\ref{rotatingmetric}), we will examine the event horizon and the ergosphere. To achieve this, we determine the two physical horizons of the theory by setting \( 1/g_{rr} = 0 \), or equivalently, \( \Delta(r) = 0 \)
\ie
r_{+} =\sqrt{M \left(M-\frac{8 \sqrt{\Theta }}{\sqrt{\pi }}\right)-a^2}+M,
\fe
which $r_{+}$ represents the outer horizon and
\ie
r_{-} = M-\sqrt{M \left(M-\frac{8 \sqrt{\Theta }}{\sqrt{\pi }}\right)-a^2},
\fe
accounts for the inner horizon instead. It is important to emphasize that as \(\Theta \to 0\), above results converge to the horizons of a standard Kerr black hole. To better illustrate the behavior of the event horizon, we present Fig. \ref{eventhor}. In the top left panel, the variation of \( r_{+} \) with \( M \) is shown for different values of \( \Theta \) with a fixed spin parameter \( a = 0.5 \). The top right panel displays how \( r_{+} \) changes with \( M \) for different values of \( a \), while keeping \( \Theta = 0.1 \) constant. The bottom panel presents a three--dimensional plot of \( r_{+} \) as a function of \( a \) and \( \Theta \), with the mass parameter fixed at \( M = 2 \). Moreover, in Fig. \ref{ergosasdasd}, it is exhibited a $3D$--dimensional depiction exhibiting both the event horizon and the ergosphere of our non--commutative black hole.

Next, we turn our attention to the investigation of the ergosphere. To do so, we focus on the condition where \( g_{tt} = 0 \), leading to the following result:
\ie
r_{e_{\pm}} =  \frac{ \pm \sqrt{2} \sqrt{-\pi  a^2 \cos (2 \theta )-\pi  a^2+2 \pi  M^2-16 \sqrt{\pi } \sqrt{\Theta } M}+2 \sqrt{\pi } M}{2 \sqrt{\pi }}.
\fe
The structure of the ergosphere is governed by the expression for \(r_{e_{\pm}}\), which depends on the spin parameter \(a\), mass \(M\), non--commutative parameter \(\Theta\), and the angular coordinate \(\theta\). Additionally, when \(\Theta \to 0\), the ergosphere reduces to that of the standard Kerr black hole. The angular dependence, particularly the \(\cos(2\theta)\) term, causes the ergosphere to deviate from a perfect sphere, appearing flattened along the poles of the rotation axis, where \(\theta = 0\) or \(\theta = \pi\). The ergosphere spans the region between the inner and outer surfaces defined by \(r_{e_{\pm}}\). Within this region, the temporal component \(g_{tt}\) becomes negative, behaving like a spatial metric component. As a consequence, particles inside the ergosphere are forced to co--rotate with the central mass in order to maintain a time--like trajectory. Any particle in this region experiences positive proper time as it moves through spacetime.

\begin{figure}
    \centering
     \includegraphics[scale=0.39]{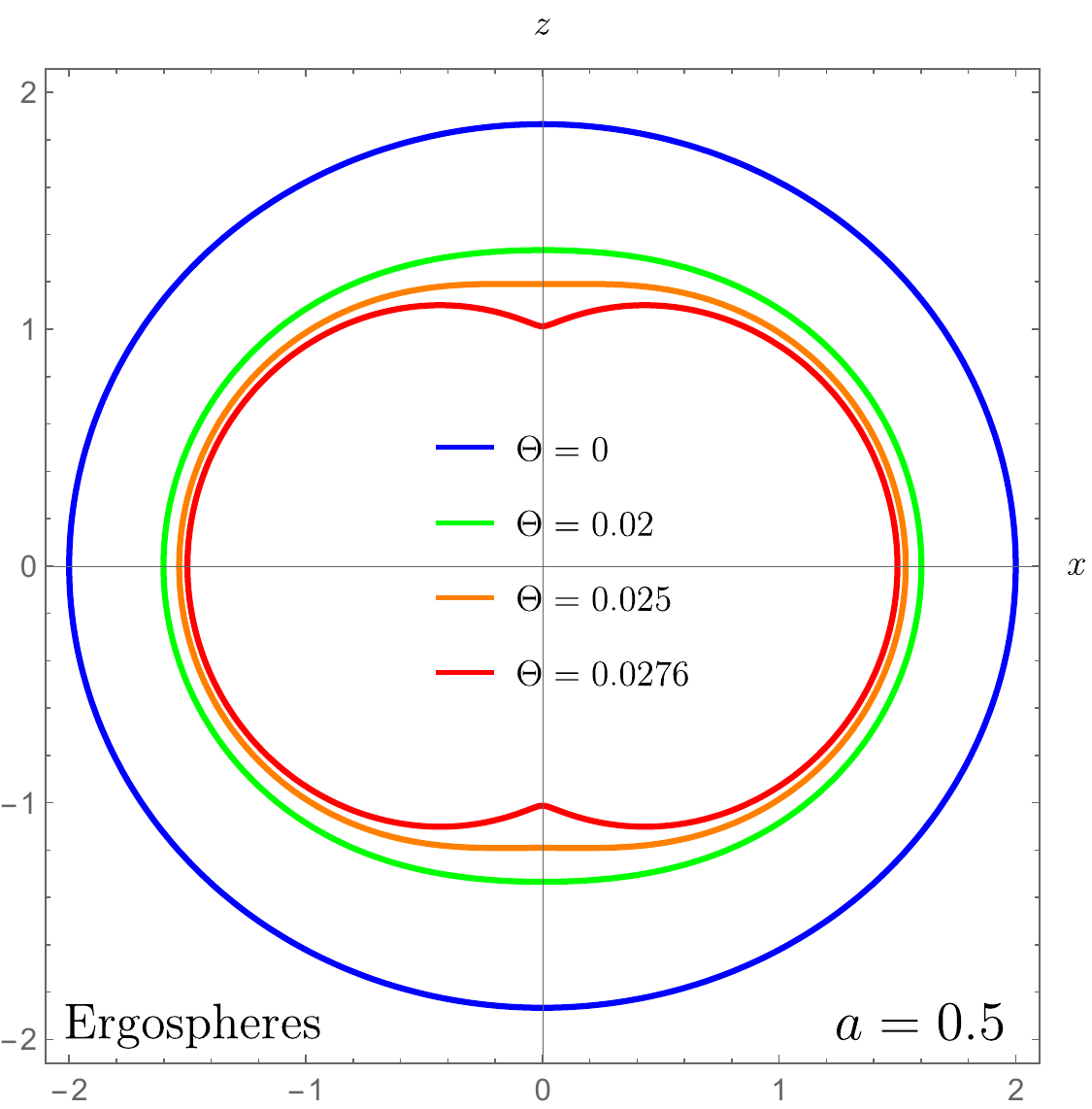}
     \includegraphics[scale=0.4]{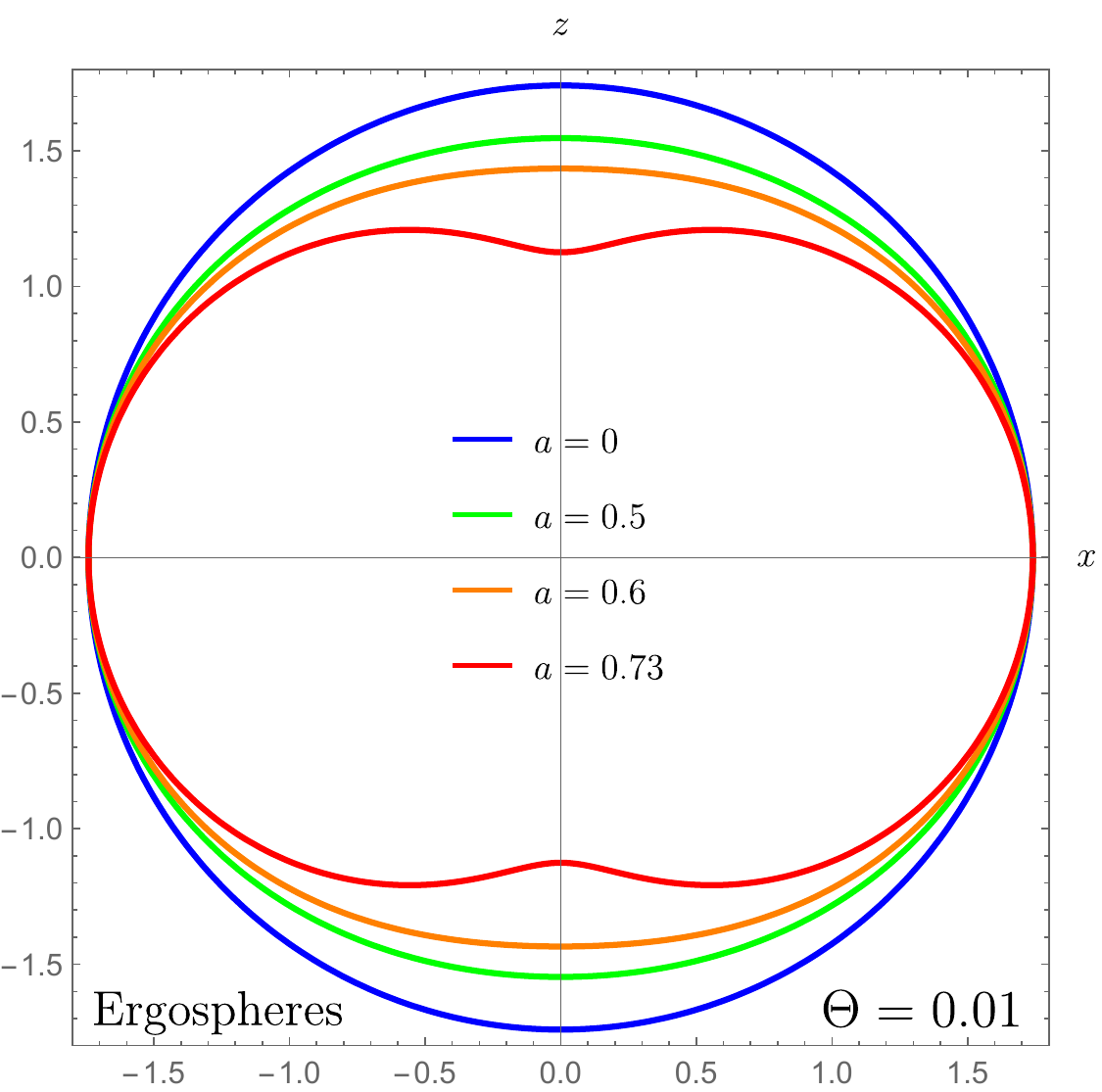}
    \caption{The parametric representation of the ergospheres for varying values of \(\Theta\) and \(a\).}
    \label{ergos}
\end{figure}

Fig. \ref{ergos} presents the parametric plot of the inner and outer ergospheres for varying values of the spin parameter \(a\), while keeping the non--commutative parameter fixed at \(\Theta = 0.01\) and mass at \(M = 1\). A key aspect to emphasize is that, compared to the standard Kerr black hole solution, non--commutativity causes a noticeable ``squeezing'' of the ergosphere along the \(z\)--axis, leading to a more compressed shape in this direction.

\subsection{Angular velocity}

\begin{figure}
    \centering
     \includegraphics[scale=0.42]{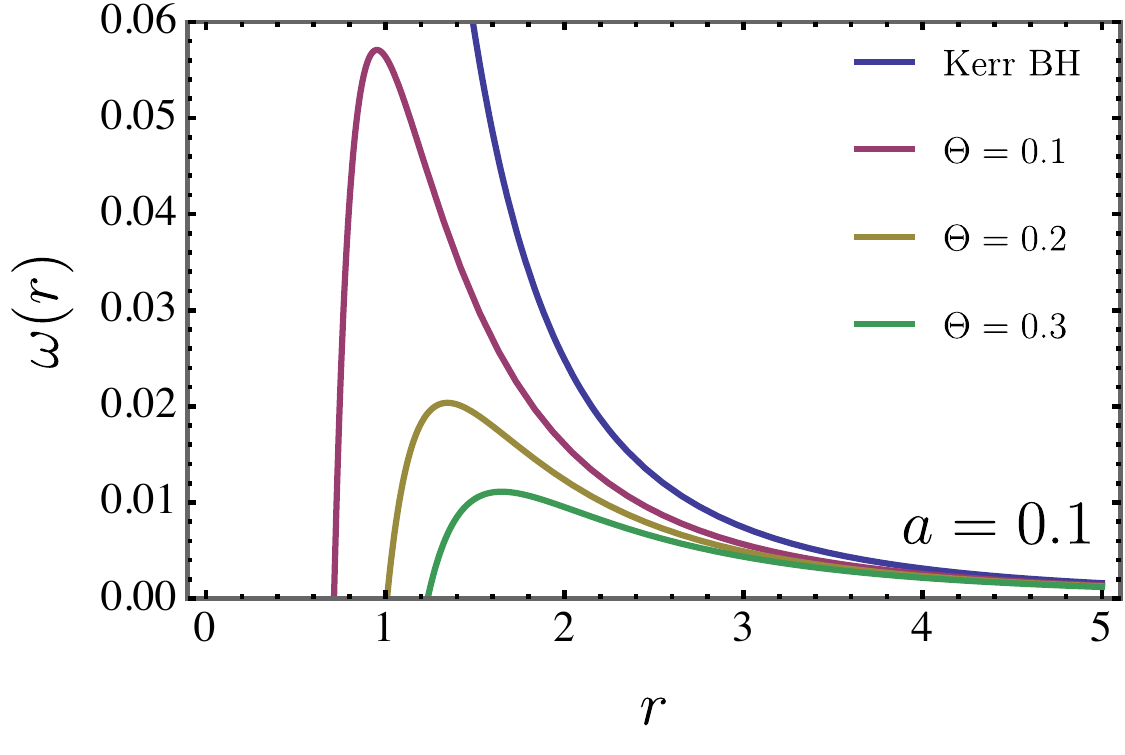}
     \includegraphics[scale=0.422]{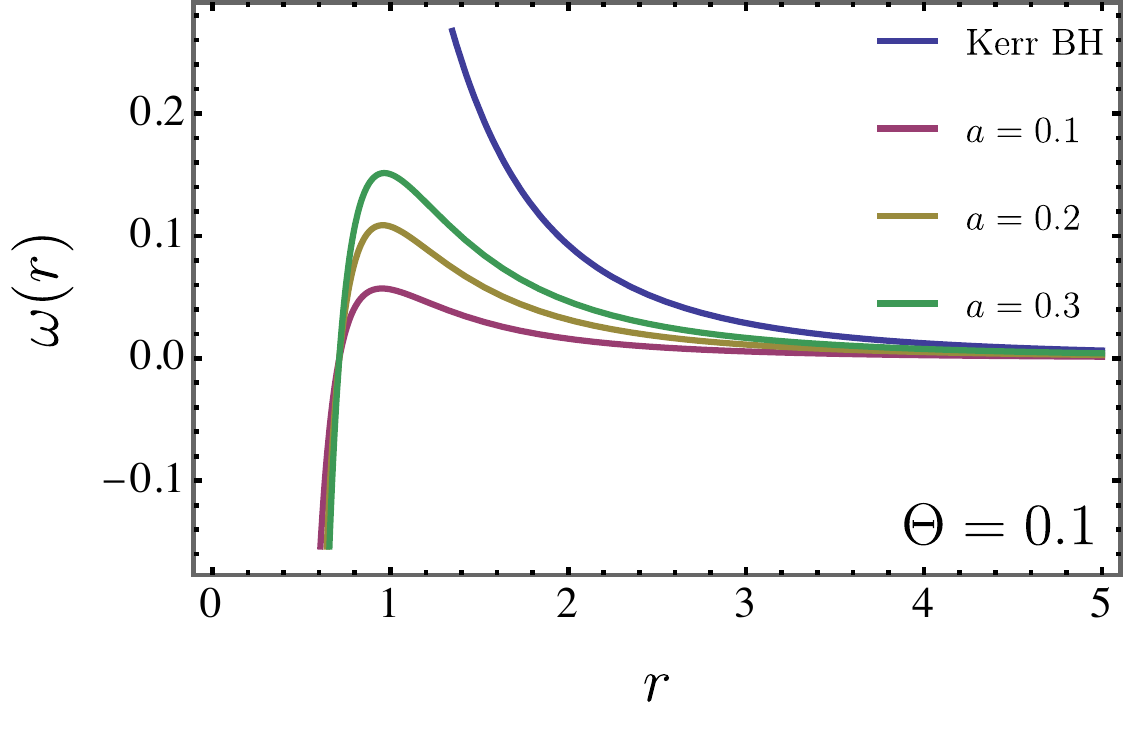}
    \caption{The angular velocity $\omega$, begin represented by two--dimensional plots for different values of $a$ and $\Theta$.}
    \label{angularvelocity2d}
\end{figure}

Similar to the Kerr black hole, our solution possesses isometries related to time translation and rotational symmetry, indicated by the existence of Killing vectors. As a result of the \textit{Lense--Thirring} effect, a stationary observer situated outside the event horizon, who has zero angular momentum with respect to an observer at infinity, will be dragged into rotational motion. In the non--commutative black hole spacetime, this frame--dragging effect arises (among other components) due to the non--zero off--diagonal metric component \(g_{t\phi}\). Accordingly, the observer's angular velocity is governed by \cite{visser2007kerr,grumiller2022black}
\begin{eqnarray}
&&\omega(r) = - \frac{g_{t\phi}}{g_{\phi\phi}} = \frac{a[a^{2} + r^{2} - \Delta(r)  ]}{(r^{2} + a^{2} )^{2} - a^{2} \Delta(r) \sin \theta } =\nonumber\\ 
&=&\frac{2 a M \left(\sqrt{\pi } r-4 \sqrt{\Theta }\right)}{\sqrt{\pi } \left(a^2+r^2\right)^2-a^2 \sin (\theta ) \left(\sqrt{\pi } \left(a^2+r (r-2 M)\right)+8 \sqrt{\Theta } M\right)}.
\end{eqnarray}
Naturally, when the non-commutative parameter \(\Theta\) is set to zero, the angular velocity reduces to the standard Kerr value, as expected. In Fig. \ref{angularvelocity2d}, we illustrate the behavior of \(\omega\) for various values of \(a\) and \(\Theta\), comparing our results with those of the Kerr solution. Overall, a decrease in \(\Theta\) leads to a reduction in \(\omega(r)\), while higher values of \(a\) correspond to an increase in the magnitude of \(\omega(r)\).

\subsection{Surface gravity}

To compute the surface gravity, we first transform the coordinates from Eq. (\ref{rotatingmetric}) into Eddington–Finkelstein coordinates \cite{christodoulou1971reversible,ruiz2019thermodynamic}, which effectively eliminate coordinate singularities at \(\Delta(r) = 0\). This process requires a redefinition of the time coordinate \(\tau\) and the angular coordinate \(\chi\), as outlined below:
\ie
\mathrm{d} \tau  = \mathrm{d}t + \frac{r^{2}+a^{2}}{\Delta(r)}\mathrm{d}r,
\fe
and
\ie
\mathrm{d}\chi = \mathrm{d} \varphi + \frac{a}{\Delta(r)} \mathrm{d}r,
\fe
resulting in the following form of the transformed metric:
\ie
\begin{split}
\mathrm{d}s^{2}  = & \frac{\Delta(r) - a^{2} \sin^{2}\theta }{\Sigma} \mathrm{d}\tau^{2} + 2 \mathrm{d}\tau \mathrm{d}r - \frac{2a(r^{2}+ a^{2} - \Delta(r) )\sin^{2}\theta  }{\Sigma} \mathrm{d}\tau\mathrm{d}\chi \\
&-2a \sin^{2}\theta \mathrm{d}r \mathrm{d}\chi - \frac{(r^{2} + a^{2} )^{2} - \Delta(r) a^{2} \sin^{2}\theta  }{\Sigma}  \mathrm{d}\chi^{2}.
\end{split}
\fe
Through this variable transformation, it becomes clear that on the hypersurfaces where \( r = r_\pm \), the Killing vectors corresponding to the coordinates can be written as:
\ie
\psi_{\pm} = \frac{\partial}{\partial \tau} + \frac{\Delta(r)}{r_{\pm}^{2} + a^{2}} \frac{\partial}{\partial \chi}.
\fe
Furthermore, the Killing vectors can be reformulated in terms of Boyer--Lindquist coordinates. Let us define a Killing horizon \( K \), characterized by the normal Killing vector \( \xi \). The surface gravity \( \kappa \) is then defined as the proportionality constant between the vectors \( \xi^\nu \nabla_\nu \xi^\mu \) and \( \xi^\mu \) \cite{wald2010general}. The expression for \( \kappa \) is given by:
\ie
\begin{split}
k_{+} = \frac{\Delta^{\prime}(r_{+})}{2(r^{2}_{+} + a^{2})}   = \frac{\sqrt{M \left(M-\frac{8 \sqrt{\Theta }}{\sqrt{\pi }}\right)-a^2}}{\left(\sqrt{M \left(M-\frac{8 \sqrt{\Theta }}{\sqrt{\pi }}\right)-a^2}+M\right)^2+a^2}.
\end{split}
\fe
As we shall explore in the following subsections, the surface gravity will be crucial in analyzing the thermodynamic quantities associated with the black hole under consideration.


\subsection{\textit{Hawking} temperature}

In 1973, Bardeen, Carter, and Hawking introduced the four laws of classical black hole mechanics, drawing an analogy to the four laws of thermodynamics \cite{bardeen1973four}. The zeroth law establishes that a black hole's surface gravity remains constant across its event horizon, comparable to the uniform temperature of a system in thermal equilibrium \cite{page2005hawking}. The first law connects changes in a black hole’s energy (mass) with variations in its area and angular momentum (and potentially its electric charge), analogous to how the first law of thermodynamics relates changes in internal energy to heat and work \cite{carlip2014black}. The second law states that the total area of a black hole’s event horizon can never decrease, reflecting the second law of thermodynamics, which holds that the entropy of an isolated system cannot decrease \cite{davies1978thermodynamics}. Finally, the third law posits that reducing a black hole’s surface gravity to zero is impossible through any physical process, similar to the third law of thermodynamics, which asserts that absolute zero cannot be attained \cite{hawking1976black}.

\begin{figure}
    \centering
     \includegraphics[scale=0.42]{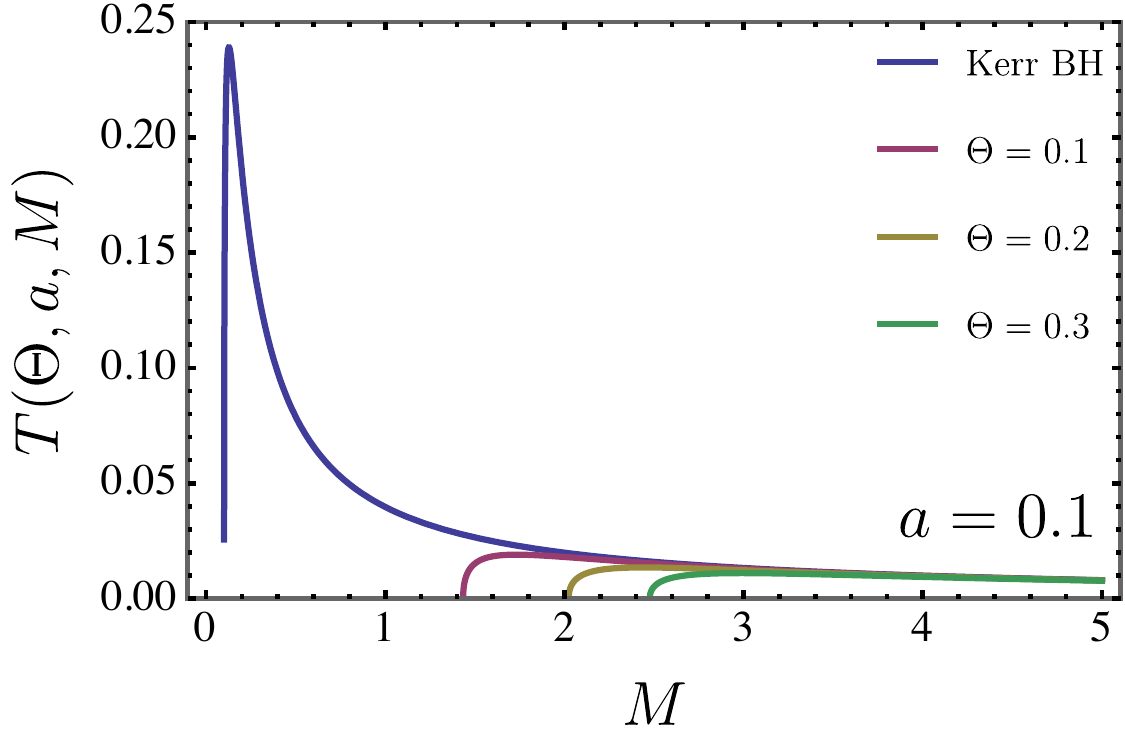}
     \includegraphics[scale=0.42]{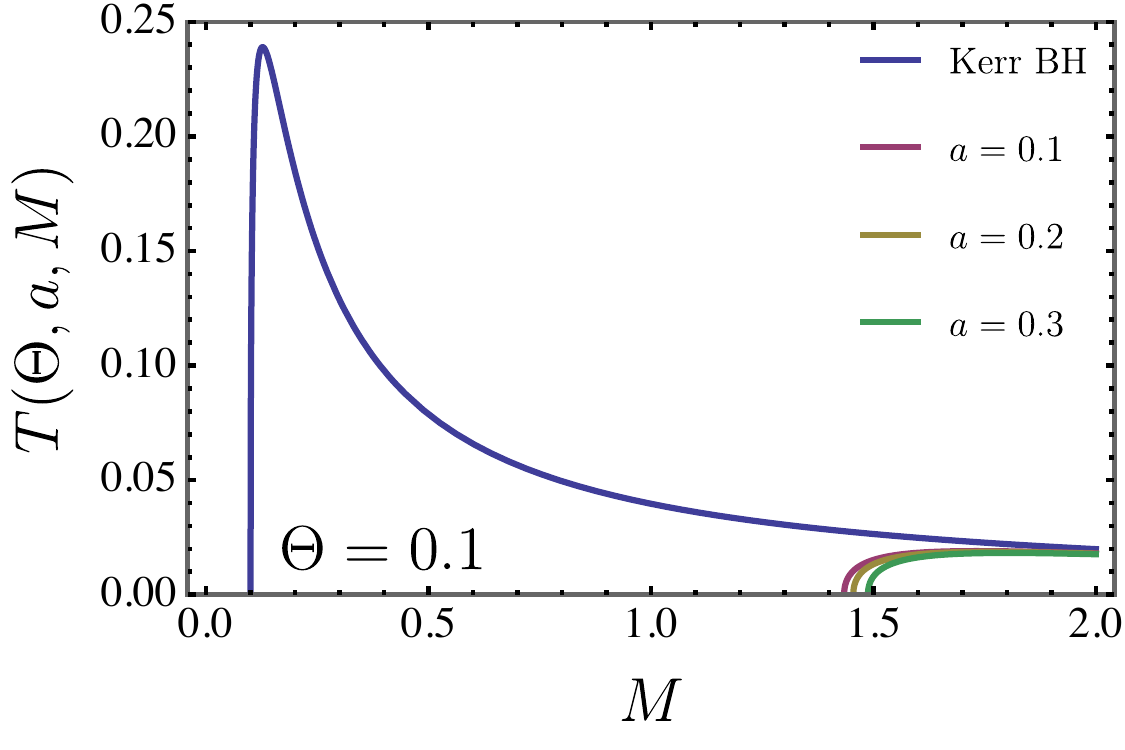}
    \caption{The \textit{Hawking} temperature as a function of $M$ for different values of $\Theta$ and a fixed value of $a=0.1$.}
    \label{Hawkingtemperature}
\end{figure}

These laws were further reinforced by the contributions of Christodoulou, who explored the irreversible processes in black hole dynamics \cite{christodoulou1970reversible}, and Bekenstein, who introduced the groundbreaking concept of black hole entropy. Bekenstein proposed that a black hole’s entropy is proportional to the area of its event horizon \cite{1bekenstein2020black,2bekenstein1974generalized}, culminating in the \textit{Bekenstein--Hawking} entropy, which bridges black hole mechanics with thermodynamic principles \cite{araujo2022thermal,furtado2023thermal}. It is important to note that, unless otherwise specified, thermodynamic quantities will be calculated assuming \(\Theta, a < M\). In addition, the \textit{Hawking} temperature is expressed as follows: 
\ie
\begin{split}
T(\Theta,a,M)  = \frac{k}{2\pi} =\frac{\sqrt{M \left(M-\frac{8 \sqrt{\Theta }}{\sqrt{\pi }}\right)-a^2}}{2 \pi  \left(\left(\sqrt{M \left(M-\frac{8 \sqrt{\Theta }}{\sqrt{\pi }}\right)-a^2}+M\right)^2+a^2\right)}.
\end{split}
\fe
This thermodynamic quantity is illustrated in Fig. \ref{Hawkingtemperature}, which demonstrates the effects of the non--commutative parameter \(\Theta\) and the rotational parameter \(a\) on the temperature. In the left panel, it is evident that increasing \(\Theta\) reduces the magnitude of \(T(\Theta, a, M)\). Similarly, a rise in \(a\) produces the same trend, leading to a further decrease in temperature. Our results are also compared with the standard Kerr black hole solution, highlighting the deviations introduced by these additional parameters.


\subsection{Entropy}

\begin{figure}
    \centering
     \includegraphics[scale=0.415]{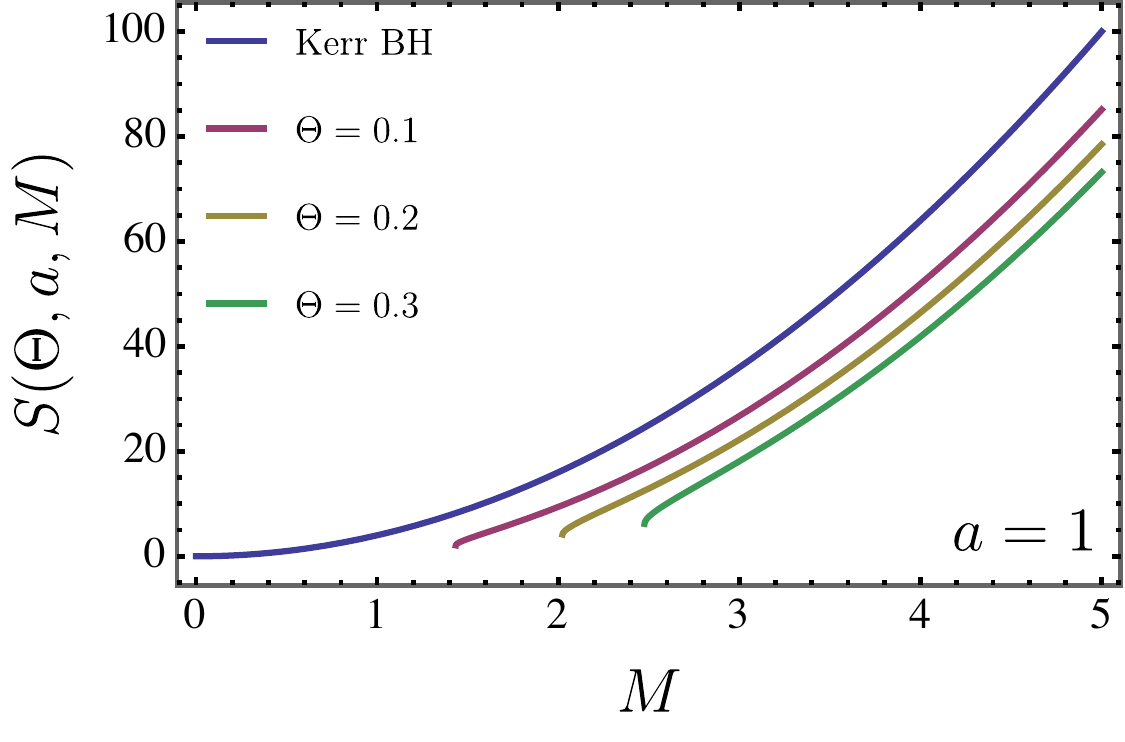}
     \includegraphics[scale=0.42]{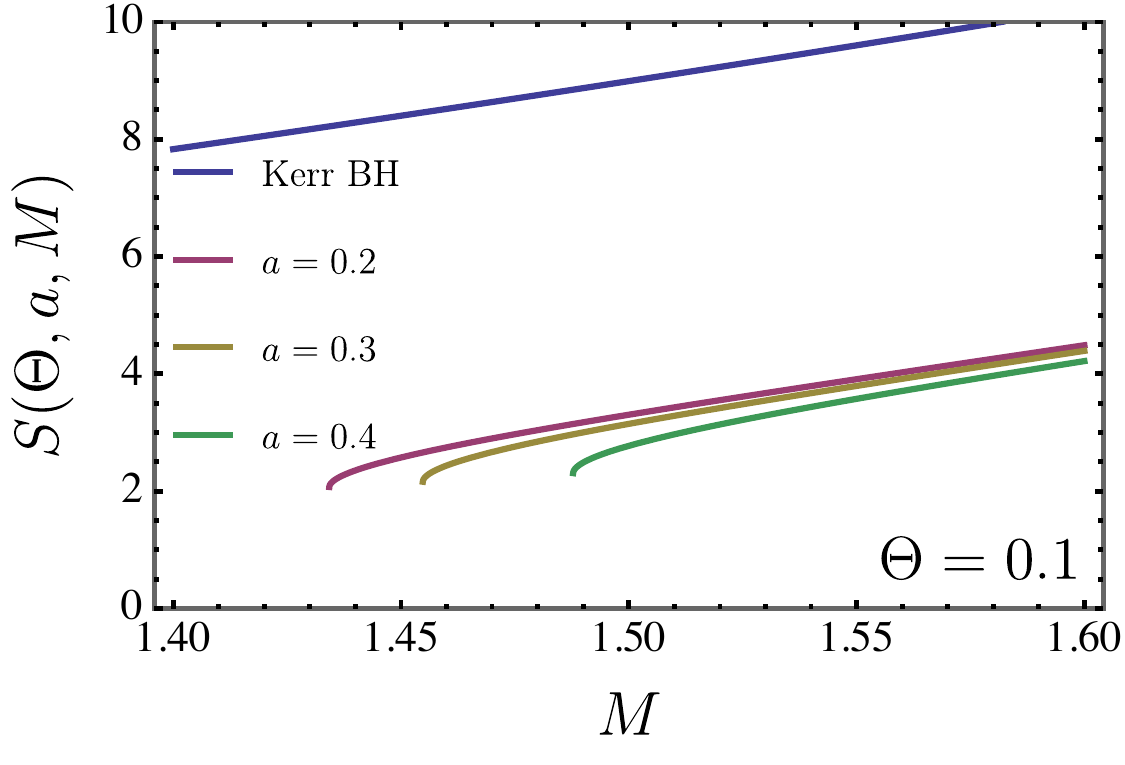}
    \caption{Entropy as a function of $M$ for various values of $\Theta$ with a fixed rotation parameter of $a=0.1$.}
    \label{entropyrot}
\end{figure}

Straightforwardly, the entropy is given by:
\ie
\begin{split}
& S(\Theta,a,M) = \, \frac{2 \sqrt{M^2 \left(2 \left(\pi ^{3/4} M \sqrt{\sqrt{\pi } \left(M^2-a^2\right)-8 \sqrt{\Theta } M} + \gamma\right)-\pi  a^2\right)}}{\sqrt{\pi }},
\end{split}
\fe
where, \(\gamma \equiv -4 \sqrt[4]{\pi } \sqrt{\Theta } \sqrt{\sqrt{\pi } \left(M^2 - a^2\right) - 8 \sqrt{\Theta } M} + 8 \Theta + \pi M^2 - 8 \sqrt{\pi } \sqrt{\Theta } M\). Fig. \ref{entropyrot} displays the entropy, illustrating the effects of the non-commutative parameter \(\Theta\) for a fixed rotational parameter \(a = 0.1\). As with the \textit{Hawking} temperature, we have compared these results with the standard Kerr black hole case, emphasizing the impact of \(\Theta\) on the system's entropy.


\subsection{Heat capacity}

Finally, for the sake of completeness, we address the study of the heat capacity as well. Thereby, we write
\ie
C_{V}(\Theta,a,M) = T\frac{\partial S}{\partial T} = T \frac{\partial S/ \partial M }{\partial T/ \partial M}  = - \frac{\Gamma}{\tilde{\Gamma}}
\fe
where 
\begin{equation}
\begin{split}
& \Gamma = 2 M^2 \left(\sqrt[4]{\pi } \sqrt{\sqrt{\pi } \left(M^2-a^2\right)-8 \sqrt{\Theta } M}-4 \sqrt{\Theta }+\sqrt{\pi } M\right) \sqrt{\sqrt{\pi } \left(M^2-a^2\right)-8 \sqrt{\Theta } M} \\
& \times \left[-\pi  a^2 \sqrt{\sqrt{\pi } \left(M^2-a^2\right)-8 \sqrt{\Theta } M}+8 \pi ^{3/4} \sqrt{\Theta } \left(a^2-5 M^2\right) \right. \\
& \left. -24 \sqrt{\Theta } M \sqrt{\pi ^{3/2} \left(M^2-a^2\right)-8 \pi  \sqrt{\Theta } M} \right. \\
& \left. +16 \Theta  \sqrt{\sqrt{\pi } \left(M^2-a^2\right)-8 \sqrt{\Theta } M}+4 \pi  M^2 \sqrt{\sqrt{\pi } \left(M^2-a^2\right)-8 \sqrt{\Theta } M} \right. \\
& \left. +\pi ^{5/4} M \left(4 M^2-3 a^2\right)+80 \sqrt[4]{\pi } \Theta  M\right]
\end{split}
\end{equation}
and
\begin{equation}
\begin{split}
\tilde{\Gamma} & = \sqrt{\pi } \left\{-\pi ^{3/4} a^2 \sqrt{\sqrt{\pi } \left(M^2-a^2\right)-8 \sqrt{\Theta } M}+4 \sqrt{\pi } \sqrt{\Theta } \left(a^2-3 M^2\right) \right. \\
& \left.-8 \sqrt[4]{\pi } \sqrt{\Theta } M \sqrt{\sqrt{\pi } \left(M^2-a^2\right)-8 \sqrt{\Theta } M} \right. \\
& \left. +\pi ^{3/4} M^2 \sqrt{\sqrt{\pi } \left(M^2-a^2\right)-8 \sqrt{\Theta } M}+\pi  M \left(M^2-2 a^2\right)+16 \Theta  M\right\} \\
& \times \left\{ M^2 \left(2 \left(\pi ^{3/4} M \sqrt{\sqrt{\pi } \left(M^2-a^2\right)-8 \sqrt{\Theta } M}  \right.\right.\right. \\
& \left.\left.\left. -4 \sqrt[4]{\pi } \sqrt{\Theta } \sqrt{\sqrt{\pi } \left(M^2-a^2\right)-8 \sqrt{\Theta } M}+8 \Theta +\pi  M^2 -8 \sqrt{\pi } \sqrt{\Theta } M\right)-\pi  a^2\right)\right\}^{1/2}.
\end{split}
\end{equation}

\begin{figure}
    \centering
     \includegraphics[scale=0.41]{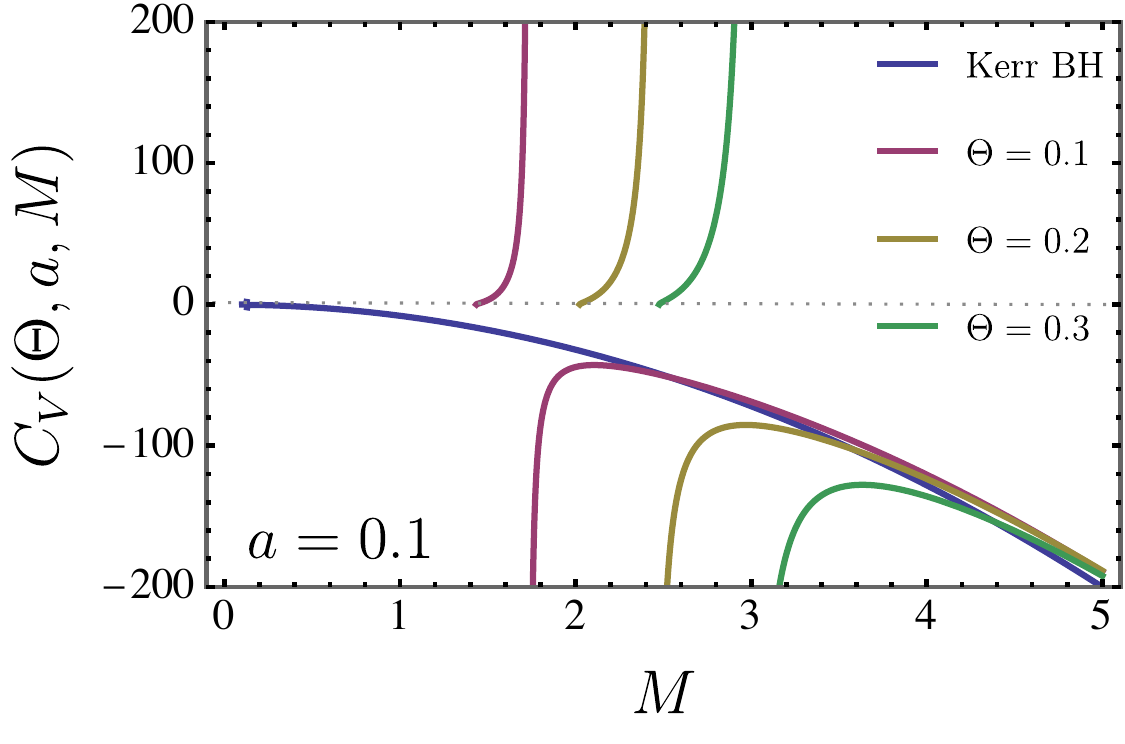}
     \includegraphics[scale=0.41]{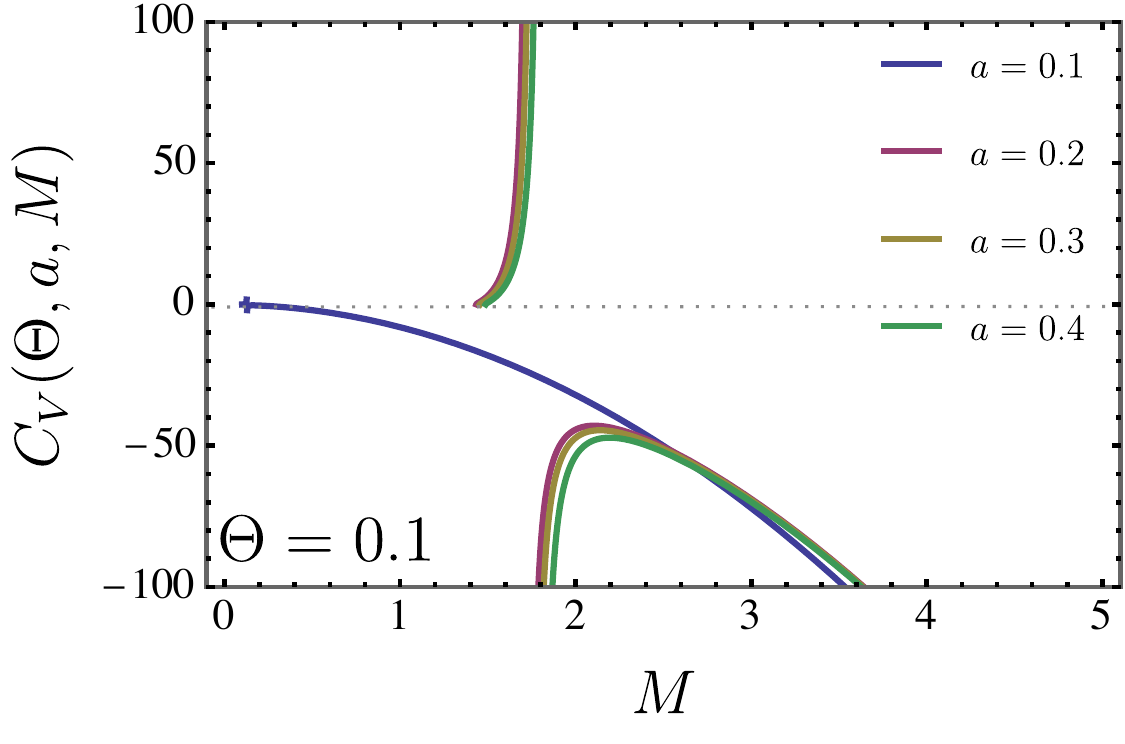}
    \caption{Heat capacity as a function of $M$ for various values of $\Theta$ and $a$.}
    \label{heatrot}
\end{figure}

Next, we examine the heat capacity in Fig. \ref{heatrot}. As before, we explore different values of \(\Theta\) while keeping the rotational parameter fixed at \(a = 0.1\). Our findings are compared with the Kerr black hole case, showing the influence of the non--commutative parameter \(\Theta\) on the heat capacity. Additionally, gravitational non--commutative theory has been investigated in various contexts, including non--commutative generalizations \cite{hrelja2024entropy}, scalar fields, Reissner--Nordström black holes \cite{ciric2018noncommutative,ciric2024noncommutative,dimitrijevic2020noncommutative}, and BTZ black holes \cite{juric2023noncommutative}.

\section{Geodesics\label{geoss}}

Our goal now is to examine how non--commutativity influences the geodesic trajectories of particles in the spacetime described by the rotating black hole. The axisymmetric nature of the metric introduces two corresponding Killing vectors, \(\partial_{t}\) and \(\partial_{\phi}\), which simplifies the analysis by allowing us to concentrate on radial geodesics. To derive the equations governing the motion of particles along these geodesics, we begin by constructing the Lagrangian, as outlined in \cite{Wald}, which will serve as the foundation for our study
\ie
\mathcal{L} = g_{\mu\nu}\dot{x}^{\mu}\dot{x}^{\nu}.
\fe
The parameter \(\mathcal{L}\) can take on values of \(-1\), \(0\), or \(1\), which correspond to timelike, null, and spacelike geodesics, respectively. In the equation mentioned earlier, the dot indicates differentiation with respect to the affine parameter \(\lambda\). We express the velocity as \(\dot{x}^{\mu} = \frac{\mathrm{d}x^{\mu}}{\mathrm{d}\lambda}\). With this velocity definition, the relevant equations can be written as follows
\ie
\begin{split}
\mathcal{L}  = & -\left[\frac{\Delta(r) -a^2\sin^2{\theta}}{\Sigma}\right]\Dot{t}^2 
   + \frac{\Sigma}{\Delta(r)}  \Dot{r}^2 + \Sigma \, \Dot{\theta}^2  \\
   &  -2a\sin^{2}\theta  \left[1 -\frac{\Delta(r) - a^{2} \sin^{2}\theta  }{\Sigma}   \right]   \Dot{t} \Dot{\phi}   +   \frac{\sin^{2}\theta}{\Sigma} \left[ (r^{2} + a^{2})^{2}  - \Delta a^{2} \sin^{2} \theta   \right]  \Dot{\phi}^2 
    \label{metricr}
    \end{split}
\fe
For the purposes of our analysis, we restrict the particle motion to the equatorial plane, setting \(\theta = \frac{\pi}{2}\). With this simplification, the equation above takes the following form
\ie
\begin{split}
\mathcal{L}  = & -\left[\frac{\Delta(r) -a^2}{\Sigma}\right]\Dot{t}^2 
   + \frac{\Sigma}{\Delta(r)}  \Dot{r}^2  \\
   &  -2a \left[1 -\frac{\Delta(r) - a^{2}   }{\Sigma}   \right]   \Dot{t} \Dot{\phi}   +   \frac{1}{\Sigma} \left[ (r^{2} + a^{2})^{2}  - \Delta a^{2}  \right]  \Dot{\phi}^2 
    \label{leriril}
    \end{split}
\fe
Given that the system preserves two key quantities, energy \(E\) and angular momentum \(L\), it follows that the equation simplifies as follows
\ie
\label{energy}
E = - g_{t\mu}\dot{x}^{\mu} =  \left(\frac{\Delta(r) - a^{2}}{\Sigma}\right) \Dot{t} + 2a \left[1 -\frac{\Delta(r) - a^{2}   }{\Sigma}   \right] \Dot{\phi},
\fe
and
\ie
\label{angularmomentum}
L = g_{\phi \mu}\dot{x}^{\mu} = - 2a \left[1 -\frac{\Delta(r) - a^{2}   }{\Sigma}   \right] \Dot{t} + \frac{1}{\Sigma} \left[ (r^{2} + a^{2})^{2}  - \Delta a^{2}  \right] \Dot{\phi}.
\fe
To simplify the process of solving Eqs. (\ref{energy}) and (\ref{angularmomentum}), we introduce a shorthand notation, defined as follows
\ie
E = A \Dot{t} + B \Dot{\phi},
\fe
and
\ie
L = - B \Dot{t} + C \Dot{\phi}, 
\fe
in which $A \equiv \frac{\Delta(r) - a^{2}}{\Sigma}$, $B \equiv a \left[1 -\frac{\Delta(r) - a^{2}   }{\Sigma}   \right]$, and $C \equiv  \frac{1}{\Sigma} \left[ (r^{2} + a^{2})^{2}  - \Delta a^{2}  \right]  $. Notice that
\ie
CE - BL = (AC + B^{2})\Dot{t} = \left[\frac{ a^2 M^2}{r^2}-\frac{f_{\Theta}(r) \left(\left(a^2+r^2\right)^2 + a^2 \Delta(r) \right)}{r^4} \right] \Dot{t} = \tilde{\Delta}(r)\Dot{t}
\fe
and
\ie
AL + BE= (AC + B^{2})\Dot{\phi} =  \left[\frac{ a^2 M^2}{r^2}-\frac{f_{\Theta}(r) \left(\left(a^2+r^2\right)^2 + a^2 \Delta(r) \right)}{r^4} \right]\Dot{\phi} = \tilde{\Delta}(r) \Dot{\phi}  .
\fe
with $ AC + B^{2} = \frac{a^2 M^2}{r^2}-\frac{f_{\Theta}(r) \left(\left(a^2+r^2\right)^2 + a^2 \Delta(r) \right)}{r^4} \equiv \tilde{\Delta}(r)$.
In this manner, we obtain
\ie
\Dot{t} = \frac{1}{\tilde{\Delta}(r) } \left[ \frac{1}{\Sigma} \left[ (r^{2} + a^{2})^{2}  - \Delta a^{2}  \right] E -  a \left[1 -\frac{\Delta(r) - a^{2}   }{\Sigma}   \right]L  \right],
\fe
\ie
\Dot{\phi} = \frac{1}{\tilde{\Delta}(r) }  \left[   \left( \frac{\Delta(r) - a^{2}}{\Sigma}   \right) L +  a \left[1 -\frac{\Delta(r) - a^{2}   }{\Sigma}   \right]   E  \right].
\fe
Next, we will derive the equation that describes the radial component of the four--velocity, expressed in terms of \(A\), \(B\), and \(C\)
\ie
\begin{split}
g_{\mu\nu}\dot{x}^{\mu}\dot{x}^{\nu} & = \mathcal{L} \\
= & - A \Dot{t}^{2} - 2 B \Dot{t} \Dot{\phi} + C \Dot{\phi}^{2} + D \Dot{r}^{2}\\
& = - [A \Dot{t} + B \Dot{\phi}] \Dot{t} + [-B \Dot{t} + C \Dot{\phi}] \Dot{\phi} + \frac{\Sigma}{\Delta(r)} \Dot{r}^{2}\\
& = - E \Dot{t} + L\Dot{\phi} + \frac{D}{\Delta(r)} \Dot{r}^{2},
\end{split}
\fe
where $D =\Sigma$. Then, the radial equation reads
\ie
\begin{split}
\Dot{r}^{2} & = \frac{\Delta(r)}{D} \left( E \Dot{t} - L \Dot{\phi} + \mathcal{L} \right) \\
& = \frac{\Delta_{eff}}{D} \left[  CE^{2} - 2 BLE -AL^{2}  + \mathcal{L} \right],
\end{split}
\fe
with $\Delta_{eff} \equiv \frac{\Delta(r)}{\tilde{\Delta}(r)}$.
Note that
\ie  
CE^{2} - 2 BLE -AL^{2}  + \mathcal{L} = \left( E - \mathcal{V}_{-} \right)\left( E + \mathcal{V}_{+} \right),
\fe
with $\mathcal{V}_{\pm} = \left(\frac{\pm \sqrt{A C L^2+B^2 L^2-C \mathcal{L}}+B L}{C}\right)$, leading to the equation below
\ie
\label{radialequation}
\Dot{r}^{2} = \frac{\Delta_{eff}}{D}\left[ \left( E - \mathcal{V}_{-} \right)\left( E + \mathcal{V}_{+} \right) \right].
\fe

In a explicit manner, ${\mathcal{V}}_{\pm}$ is written
\ie
\begin{split}
\mathcal{V}_{\pm} & = \frac{1}{\sqrt{\pi } r \left(a^2 (2 M+r)+r^3\right)-8 a^2 \sqrt{\Theta } M} \\
& \times  \left[ \pm \sqrt[4]{\pi } \Sigma  \left(\frac{L^2 r^2 \left(2 a^2+r^2\right) \left(8 \sqrt{\Theta } M+\sqrt{\pi } r (r-2 M)\right)}{\Sigma ^2}+\sqrt{\pi } a^2 L^2 \right.\right. \\
& \left.\left. + \frac{8 a^2 \sqrt{\Theta } M \left(\mathcal{L}-2 L^2\right)-\sqrt{\pi } r \left(a^2 \left(2 L^2 (r-2 M)+\mathcal{L} (2 M+r)\right)+r^3 \mathcal{L}\right)}{\Sigma }\right)^{1/2} \right. \\
& \left. -8 a \sqrt{\Theta } L M + \sqrt{\pi } a L \left(2 M r-r^2+\Sigma \right) \right].
\end{split}
\fe

\begin{figure}
    \centering
     \includegraphics[scale=0.4]{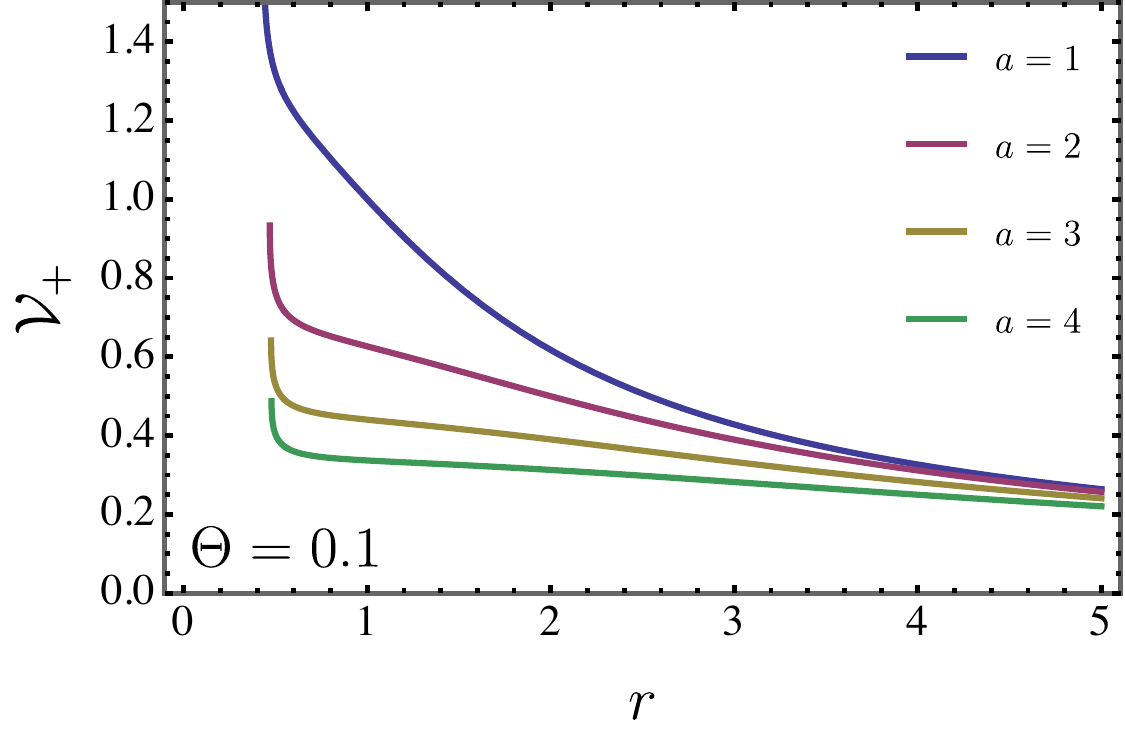}
    \includegraphics[scale=0.4]{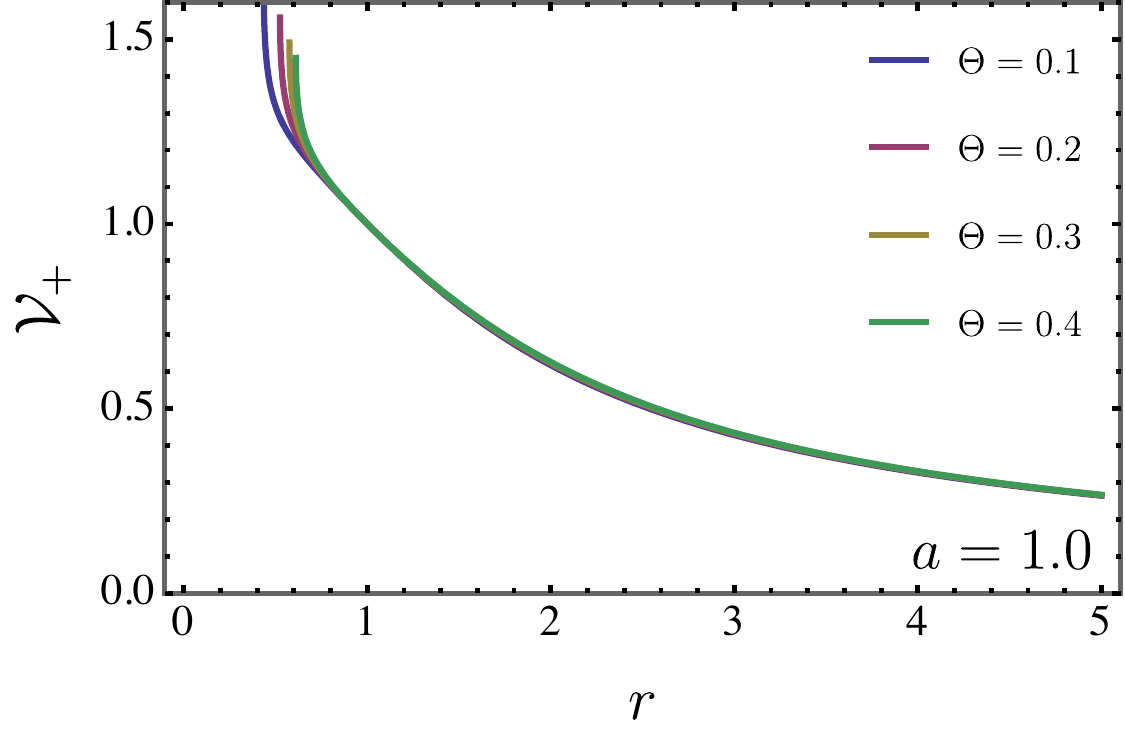}
    \caption{Potential $\mathcal{V}_{+}$ is represented for different configurations of $a$ and $\Theta$.}
    \label{potentialsplus}
\end{figure}

\begin{figure}
    \centering
     \includegraphics[scale=0.4]{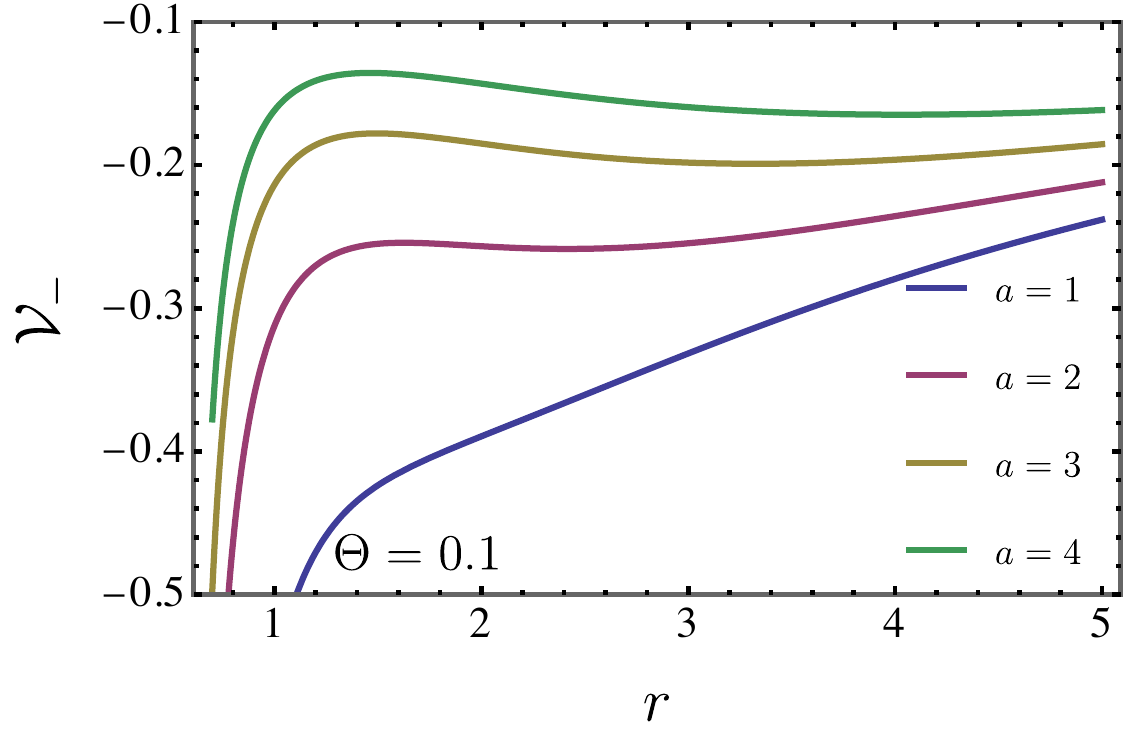}
    \includegraphics[scale=0.4]{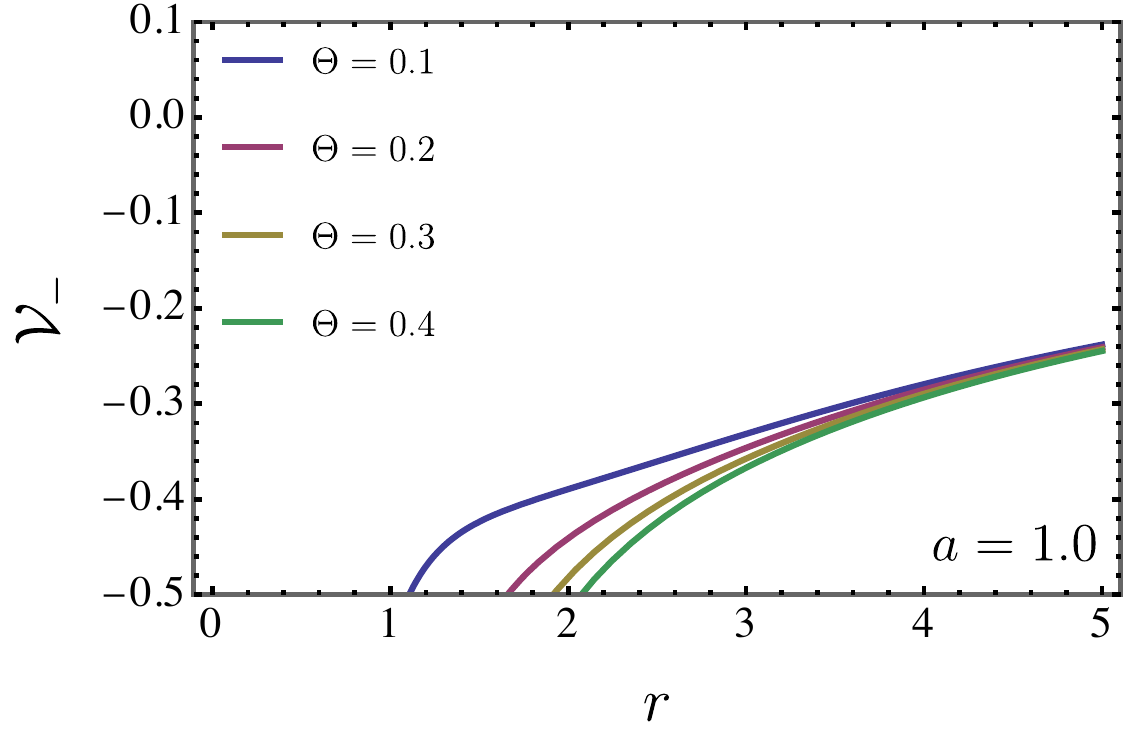}
    \caption{Potential $\mathcal{V}_{-}$ is represented for for different configurations of $a$ and $\Theta$.}
    \label{potentialsminus}
\end{figure}

\begin{figure}
    \centering
     \includegraphics[scale=0.6]{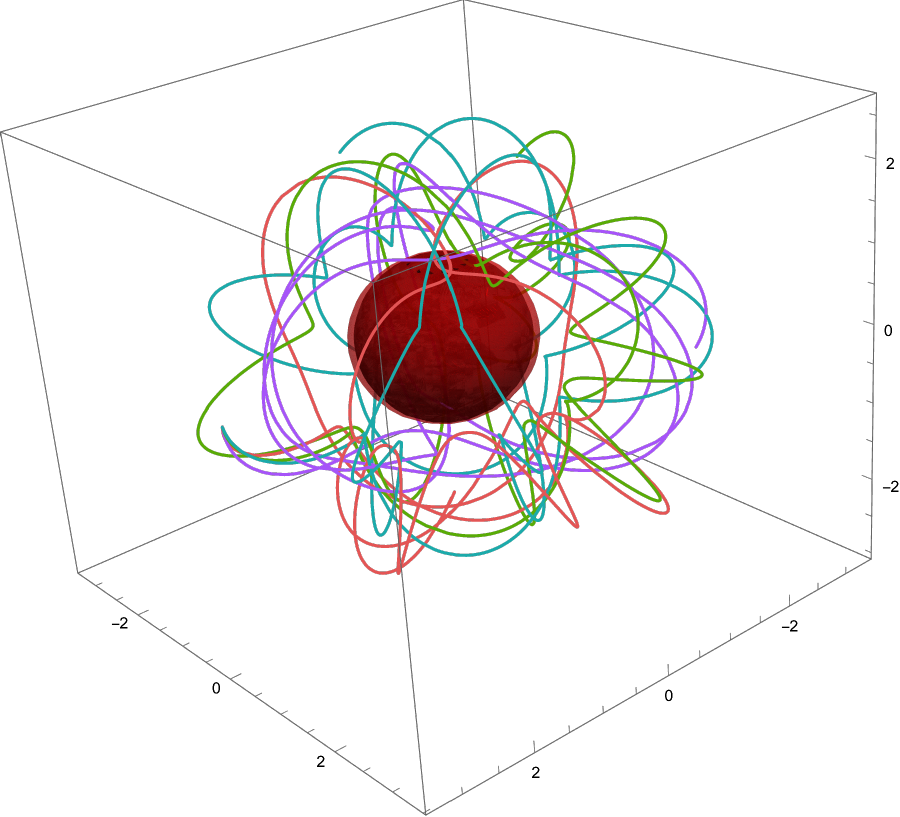}
    \caption{A 3D geodesic representation is shown, where the red ellipsoid depicts the ergosphere and the black ellipsoid illustrates the event horizon. Each colored curve corresponds to a geodesic obtained by numerical solving the equations of motion with a distinct set of initial conditions for \((t,r,\theta,\phi)\) and their respective derivatives \(\dot{t},\dot{r},\dot{\theta},\dot{\phi}\).}
    \label{geodesicsfull}
\end{figure}

To explore the behavior of the potentials \(\mathcal{V}_{\pm}\), Figs. \ref{potentialsplus} and \ref{potentialsminus} illustrate the timelike case with \(\mathcal{L} = -1\). In Fig. \ref{potentialsplus}, the dependence of \(\mathcal{V}_{+}\) on the radial coordinate \(r\) is presented. The left panel shows how the potential changes for different values of \(a\), while \(\Theta\) is kept constant. On the right, the effect of varying \(\Theta\) is examined with a fixed \(a\). Similarly, Fig. \ref{potentialsminus} displays the behavior of \(\mathcal{V}_{-}\), applying the same parameters for both \(\Theta\) and \(a\), allowing for a direct comparison between the two potentials.
Furthermore, for a better interpretation of the geodesics, we display \ref{geodesicsfull}.

\subsection{The radial acceleration analysis for null geodesics}

\begin{figure}
    \centering
     \includegraphics[scale=0.4]{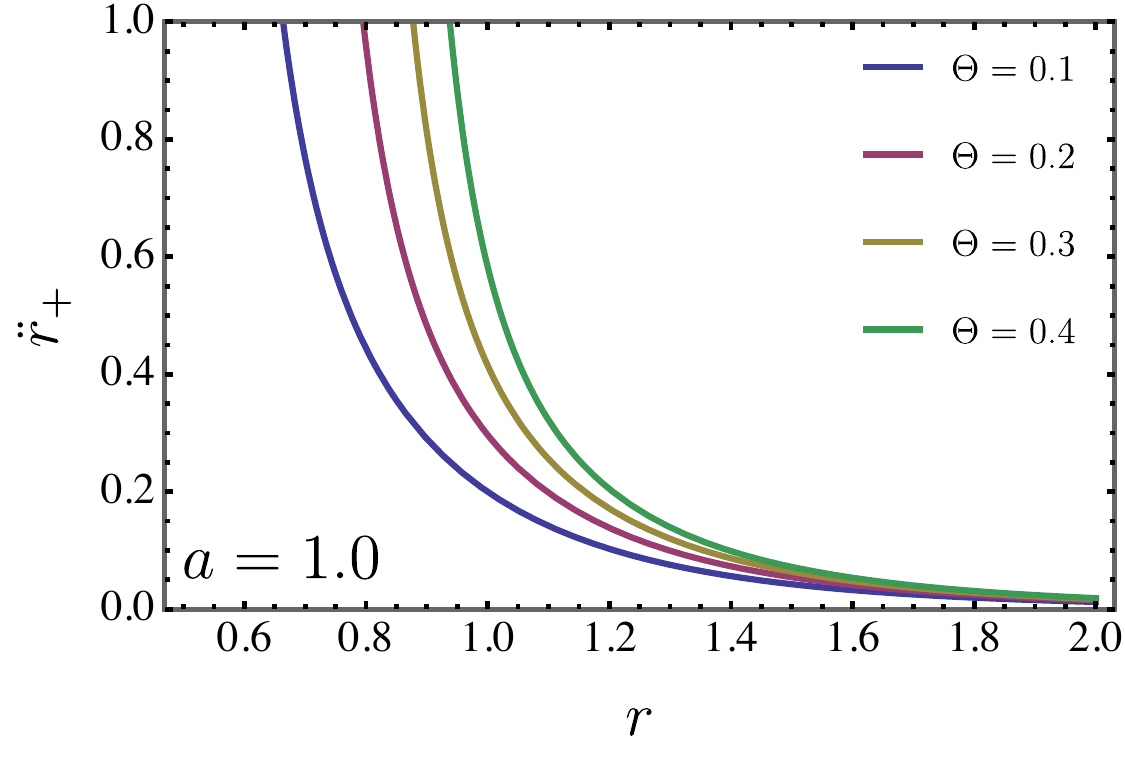}
    \includegraphics[scale=0.4]{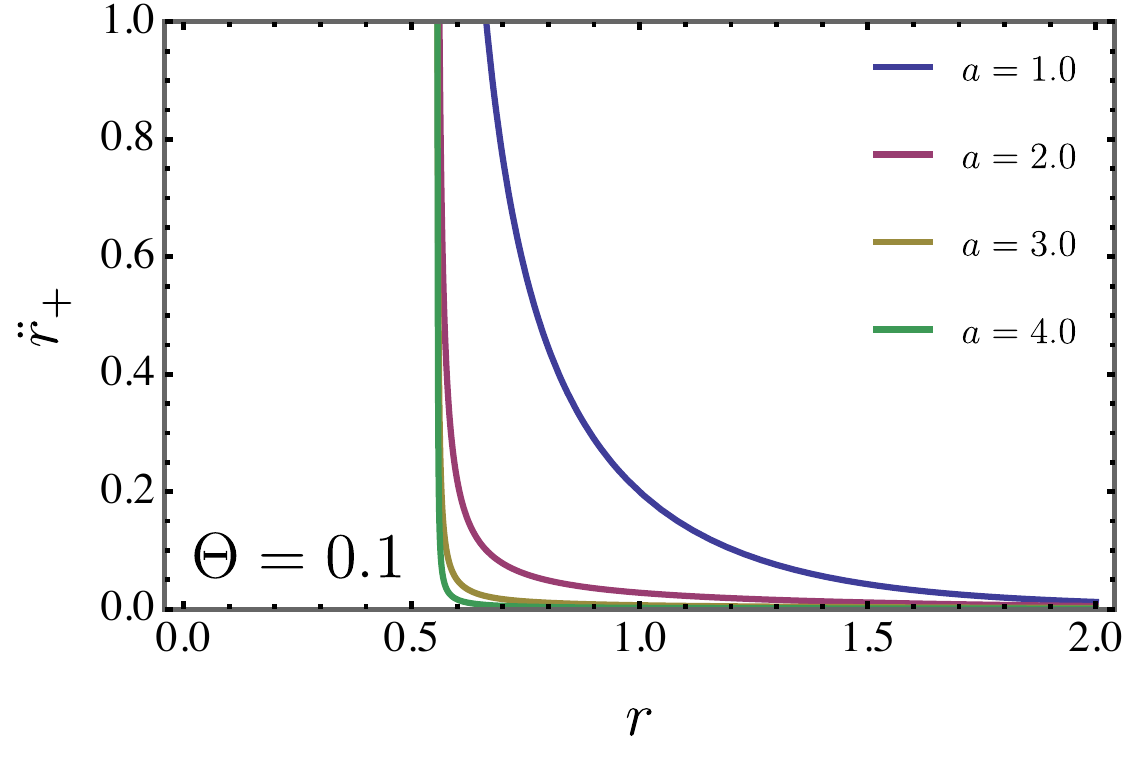}
    \caption{The acceleration \(\Ddot{r}_{+}\) is depicted as a function of \(r\) across different values of the parameters \(a\) and \(\Theta\).}
    \label{rplusa}
\end{figure}

\begin{figure}
    \centering
     \includegraphics[scale=0.4]{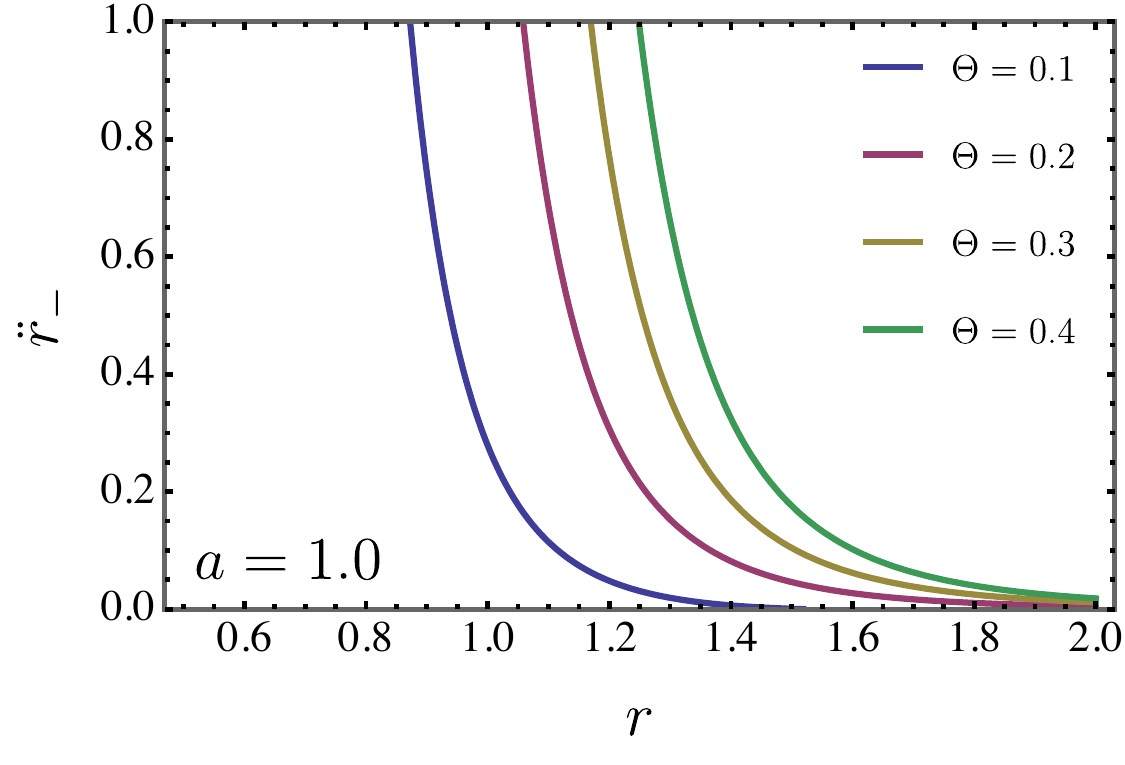}
    \includegraphics[scale=0.4]{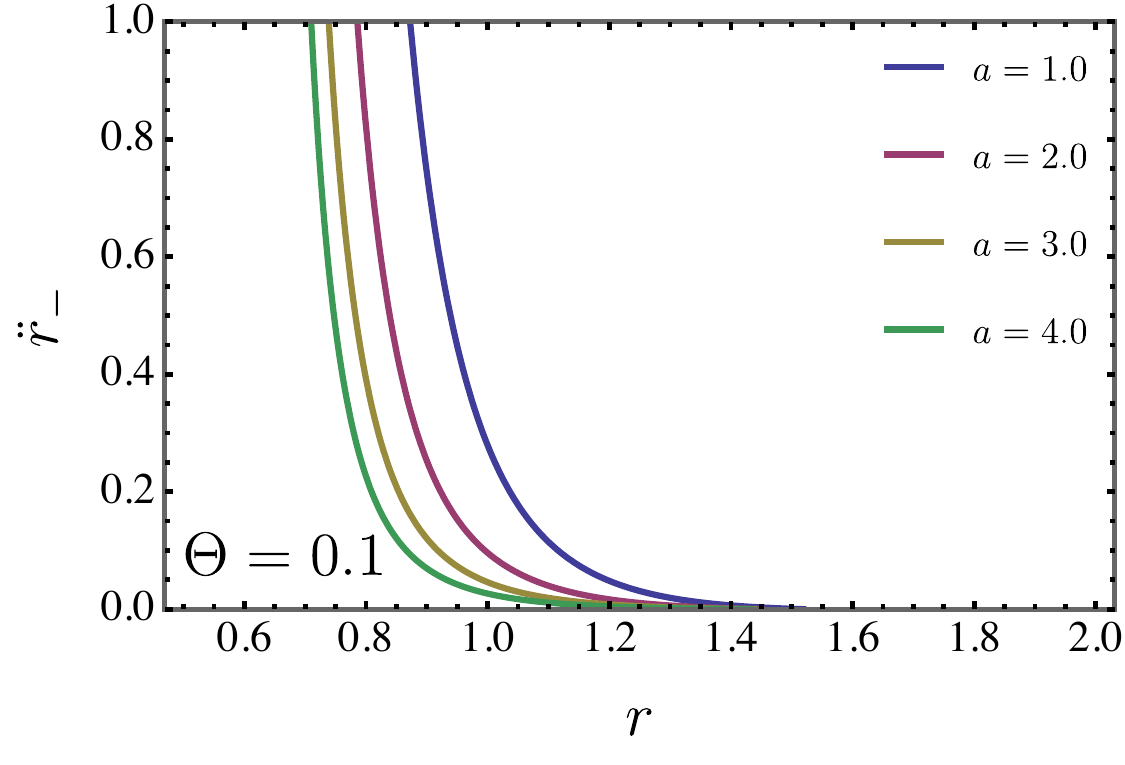}
    \caption{The acceleration \(\Ddot{r}_{-}\) is plotted as a function of \(r\) for various configurations of the parameters \(a\) and \(\Theta\).}
    \label{rminusa}
\end{figure}

For the case of null geodesics, the radial equation (\ref{radialequation}) takes the following form
\ie
\begin{split}
\Dot{r}^{2} & =  \frac{\Delta_{eff} }{D} \left( E - \mathcal{V}_{-} \right)\left( E + \mathcal{V}_{+} \right)  \\
& = \frac{\Sigma  \left(\sqrt{\pi } \left(a^2+r (r-2 M)\right)+8 \sqrt{\Theta } M\right) \times \left( E - \mathcal{V}_{-} \right)\left( E + \mathcal{V}_{+} \right)}{r^2 \left(2 a^2+r^2\right) \left(8 \sqrt{\Theta } M+\sqrt{\pi } r (r-2 M)\right)-2 a^2 \Sigma  \left(8 \sqrt{\Theta } M+\sqrt{\pi } r (r-2 M)\right)+\sqrt{\pi } a^2 \Sigma ^2} .
\end{split}
\fe
For null geodesics to be physically valid, the condition \(\Dot{r}^2 \geq 0\) must be satisfied. By examining the expression, it becomes clear that massless particles can follow null geodesics when the inequality \(\frac{\Sigma  \left(\sqrt{\pi } \left(a^2+r (r-2 M)\right)+8 \sqrt{\Theta } M\right) }{r^2 \left(2 a^2+r^2\right) \left(8 \sqrt{\Theta } M+\sqrt{\pi } r (r-2 M)\right)-2 a^2 \Sigma  \left(8 \sqrt{\Theta } M+\sqrt{\pi } r (r-2 M)\right)+\sqrt{\pi } a^2 \Sigma ^2}\) holds true. This inequality is fulfilled when the energy constant \(E\) lies outside the range defined by the potentials, ensuring the geodesic path is allowed. Thus, null geodesics are permitted when the following condition is met
\[
E < \mathcal{V}_{-} \quad \text{or} \quad E > \mathcal{V}_{+}.
\]
As a result, the region where \(\mathcal{V}_{-} < E < \mathcal{V}_{+}\) is inaccessible for particle trajectories. To gain deeper insight into the orbital dynamics, it is crucial to determine the radial acceleration. This can be done by differentiating equation (\ref{radialequation}) with respect to the affine parameter \(s\), leading to the following expression
\ie
\begin{split}
2 \dot{r} \ddot{r} =  \left[ \left( \frac{\Delta_{eff}}{D} \right)^{\prime} (E- \mathcal{V}_{+})(E-\mathcal{V}_{-}) - \frac{\Delta_{eff} }{D}\mathcal{V}_{+}^{\prime}(E-\mathcal{V}_{-})\right. \left.  - \frac{\Delta_{eff} }{D}\mathcal{V}_{-}^{\prime}(E-\mathcal{V}_{+}) \right] \Dot{r},
\end{split}
\fe
or
\ie
\begin{split}
\Ddot{r} =  \frac{1}{2}\left( \frac{\Delta_{eff}}{D} \right)^{\prime} (E- \mathcal{V}_{+})(E-\mathcal{V}_{-}) - \frac{\Delta_{eff}}{2 D} \left[ \mathcal{V}_{+}^{\prime}(E-\mathcal{V}_{-}) - \mathcal{V}_{-}^{\prime}(E-\mathcal{V}_{+}) \right].
\end{split}
\fe
In this scenario, differentiation with respect to the radial coordinate \(r\) is indicated by the prime symbol (\('\)). We now explore the analysis of radial acceleration at the points where the radial velocity \(\dot{r}\) becomes zero. This occurs when the energy parameter \(E\) aligns with either of the potential values, \(\mathcal{V}_{+}\) or \(\mathcal{V}_{-}\)
\ie
 \Ddot{r}_{+} = - \frac{\Delta_{eff}}{2 D}  \mathcal{V}_{+}^{\prime}(\mathcal{V}_{+} -\mathcal{V}_{-}), \,\,\,\,\,\,\text{if} \,\,\,\,\, E = \mathcal{V}_{+},
\fe
and
\ie
 \Ddot{r}_{-} = - \frac{\Delta_{eff}}{2 D}  \mathcal{V}_{-}^{\prime}(\mathcal{V}_{-}-\mathcal{V}_{+}), \,\,\,\,\,\,\text{if} \,\,\,\,\, E = \mathcal{V}_{-},
\fe
with \(r_{\pm}\) representing the radial accelerations. To enhance the reader's understanding of the quantity \(\Ddot{r}_{\pm}\), we present Figs. \ref{rplusa} and \ref{rminusa} for reference.

\section{Shadows \label{shaaa}}

\begin{figure}
    \centering
    \includegraphics[scale=0.45]{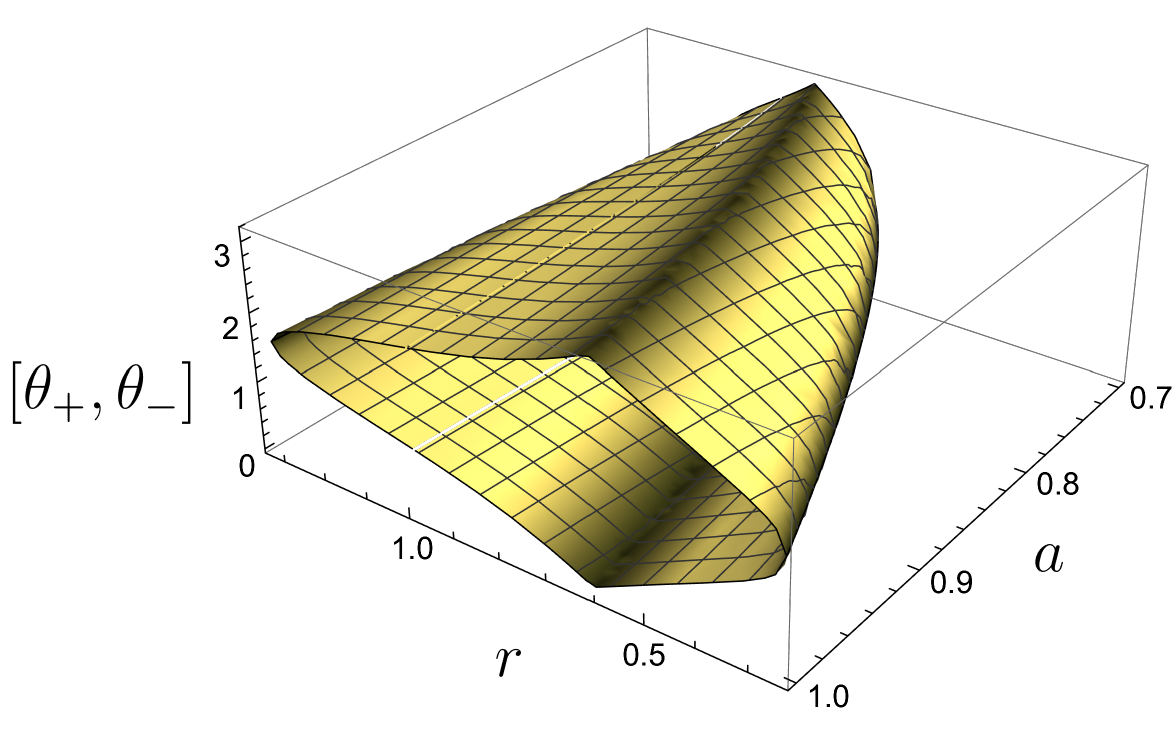}
    \caption{Parameters $\theta_{+}$ and $\theta_{-}$ are shown for different $a$ and for a fixed value of $\Theta =0.1$.}
    \label{thetabehavior}
\end{figure}

The behavior of photon geodesics is entirely determined by the geometry and symmetries of the spacetime. Therefore, analyzing null geodesics reveals important aspects of the modified gravity theory. For photon motion in this spacetime, four constants of motion play a key role: the rest mass \(m = 0\), the energy \(E\), given by \(E = -g_{t\phi}\dot{\phi} - g_{tt}\dot{t}\), and the azimuthal angular momentum \(L_z = g_{\phi\phi}\dot{\phi} + g_{\phi t}\dot{t}\). In addition, the Carter constant \(Q\), which arises from an additional hidden symmetry, enables the formulation of first-order equations for the null geodesics. These equations can be derived through Carter's approach using the Hamilton--Jacobi equation, as shown below
\ie
\begin{split}
& \Sigma \dot{t} = \frac{r^{2}+a^{2}}{\Delta} \left[ E(r^{2} + a^{2}) - a L_{z}   \right] - a(aE\sin^{2}\theta - L_{z}), \\
& \Sigma \dot{\phi} = \frac{a}{\Delta} [E(r^{2}+a^{2}) - a L_{z}] - \left( aE - \frac{L_{z}}{\sin^{2}\theta} \right), \\
& \Sigma^{2} \dot{r}^{2} = \left[ (r^{2} + a^{2})  - a L_{z}  \right]^{2} - \Delta(r) \left[ \mathcal{K} + (aE - L_{z})^{2}  \right] \equiv \mathcal{R}(r), \\
&\Sigma^{2} \dot{\theta}^{2} = \mathcal{K} - \left(  \frac{L_{z}^{2}}{\sin^{2}\theta} -a^{2}E^{2}   \right) \cos^{2}\theta \equiv \Tilde{\Theta}(r).
\end{split}
\fe
We introduce two re--scaled energy parameters, \(\xi = L/E\) and \(\eta = K/E^2\), which will be referred to as the critical impact parameters within the functions \(R(r)\) and \(\Tilde{\Theta}(r)\). Our analysis focuses on the region outside the event horizon, specifically on spherical photon orbits--null geodesics confined to specific radii \(r_p\). These orbits are characterized by the conditions \(\mathcal{R}'(r_p) = 0\) and \(\mathcal{R}''(r_p) = 0\) for all \(r_p > r_+\) \cite{afrim}. This approach allows us to calculate the critical impact parameters for the non--commutative black hole under consideration. Thus, we express the equations as follows
\ie
\begin{split}
& \xi_{c} = \frac{(a^{2}+r^{2})\Delta^{\prime}(r) - 4 r \Delta(r) }{a \Delta^{\prime}(r)},\\
& \eta_{c}   = \frac{r^{2}(  8 \Delta(r) (2 a^{2} + r\Delta^{\prime} (r)   )   -r^{2} \Delta^{\prime 2}(r) - 16\Delta^{2}(r) )  }{a^{2}\Delta^{\prime 2}(r)}.
\end{split}
\fe
The photon orbits that are spherical are constrained within a three--dimensional area known as the photon shell, which is essential for defining the bright boundary of the shadow's silhouette. This region is characterized by several parameters: time \(t\), which extends from \(-\infty\) to \(\infty\); radial position \(r_p\), spanning between \(r_{p-}\) and \(r_{p+}\); azimuthal angle \(\phi\), which ranges from 0 to \(2\pi\); and polar angle \(\theta\), varying between \(\theta_{-}\) and \(\theta_{+}\), where \(\theta_{\mp} = \arccos(\mp\sqrt{\tilde{\omega}})\)
\ie
\tilde{\omega} = \frac{a^{2}-\eta_{c} - \xi^{2}_{c} + \sqrt{(a^{2} - \eta_{c} - \xi_{c}^{2})^{2} + 4 a^{2}\eta_{c}  } }{2 a^{2}}.
\fe
To gain a clearer understanding of our results, we present the three-dimensional behavior of \(\theta_{+}\), \(\theta_{-}\), and the entire range \(\theta_{+} + \theta_{-}\) (see Fig. \ref{thetabehavior}).
It is important to note that \(r_{p\mp}\) represent the radii of the prograde and retrograde photon orbits, respectively, which are determined by solving \(\eta_c = 0\) with the condition \(\xi_c(r_{p\mp}) \gtrless 0\). Specifically, the value of \(r_{p}\) is given by the following expression
\ie
\begin{split}
r_{p} = & \frac{1}{2} \sqrt{\psi -\frac{-\frac{8 \left(4 \pi  a^2 M-96 \sqrt{\pi } \sqrt{\Theta } M^2\right)}{\pi }-6 M \left(\frac{4 \left(9 \pi  M^2+32 \sqrt{\pi } \sqrt{\Theta } M\right)}{\pi }-36 M^2\right)}{4 \sqrt{\rho }}} +\frac{3 M}{2} \\
& -\frac{1}{2} \sqrt{\frac{24 a^2 M^2}{\sqrt[3]{\zeta }}-\frac{128 a^2 \sqrt{\Theta } M}{\sqrt{\pi } \sqrt[3]{\zeta }}+\frac{\sqrt[3]{\zeta }}{3}+\frac{27 M^4}{\sqrt[3]{\zeta }}-\frac{384 \sqrt{\Theta } M^3}{\sqrt{\pi } \sqrt[3]{\zeta }} +\frac{4096 \Theta  M^2}{3 \pi  \sqrt[3]{\zeta }}+3 M^2-\frac{64 \sqrt{\Theta } M}{3 \sqrt{\pi }}}
\end{split},
\fe
where
\ie
\begin{split}
\nonumber
\zeta = & 216 a^4 M^2+972 a^2 M^4-\frac{12096 a^2 \sqrt{\Theta } M^3}{\sqrt{\pi }}+\frac{36864 a^2 \Theta  M^2}{\pi } +729 M^6 \\
& -\frac{15552 \sqrt{\Theta } M^5}{\sqrt{\pi }}+\frac{110592 \Theta  M^4}{\pi }-\frac{262144 \Theta ^{3/2} M^3}{\pi ^{3/2}} \\
& +\frac{1}{\pi^{3/2}} \times 8 \left[729 \pi ^3 a^8 M^4+729 \pi ^3 a^6 M^6+11664 \pi ^{5/2} a^6 \sqrt{\Theta } M^5  \right. \\
& \left. - 248832 \pi ^2 a^6 \Theta  M^4+884736 \pi ^{3/2} a^6 \Theta ^{3/2} M^3+17496 \pi ^{5/2} a^4 \sqrt{\Theta } M^7 \right. \\
& \left.
-388800 \pi ^2 a^4 \Theta  M^6+2875392 \pi ^{3/2} a^4 \Theta ^{3/2} M^5-7077888 \pi  a^4 \Theta ^2 M^4\right]^{1/2},
\end{split}
\fe
\ie
\begin{split}
\nonumber
\psi & =-\frac{24 a^2 M^2}{\sqrt[3]{\zeta }}+\frac{128 a^2 \sqrt{\Theta } M}{\sqrt{\pi } \sqrt[3]{\zeta }}-\frac{\sqrt[3]{\zeta }}{3}-\frac{27 M^4}{\sqrt[3]{\zeta }}+\frac{384 \sqrt{\Theta } M^3}{\sqrt{\pi } \sqrt[3]{\zeta }}\\
& -\frac{4096 \Theta  M^2}{3 \pi  \sqrt[3]{\zeta }}-\frac{2 \left(9 \pi  M^2+32 \sqrt{\pi } \sqrt{\Theta } M\right)}{\pi }+24 M^2+\frac{64 \sqrt{\Theta } M}{3 \sqrt{\pi }},
\end{split}
\fe
\ie
\nonumber
\rho =\frac{24 a^2 M^2}{\sqrt[3]{2 \zeta }}-\frac{128 a^2 \sqrt{\Theta } M}{\sqrt{\pi } \sqrt[3]{\zeta }}+\frac{\sqrt[3]{\zeta }}{3}+\frac{27 M^4}{\sqrt[3]{2 \zeta }}-\frac{384 \sqrt{\Theta } M^3}{\sqrt{\pi } \sqrt[3]{2 \zeta }}+\frac{4096 \Theta  M^2}{3 \pi  \sqrt[3]{\zeta }}+3 M^2-\frac{64 \sqrt{\Theta } M}{3 \sqrt{\pi }}.
\fe

The transition point for photons between prograde and retrograde orbits occurs at the intermediate radius \(r_{p0}\), obtained by solving the condition \(\xi_c = 0\). This specific radius defines spherical photon orbits with zero angular momentum, and it can be written as 
\ie
\begin{split}
r_{p 0} =  & \frac{\sqrt[3]{2} \left(3 \pi  a^2+3 \pi  M^2-16 \sqrt{\pi } \sqrt{\Theta } M\right)}{\sqrt{\pi } \sqrt[3]{54 \pi ^{3/2} a^2 M+\lambda +54 \pi ^{3/2} M^3-432 \pi  \sqrt{\Theta } M^2}}+M \\
& +\frac{1}{3 \sqrt[3]{2} \sqrt{\pi }} \sqrt[3]{54 \pi ^{3/2} a^2 M+\lambda +54 \pi ^{3/2} M^3-432 \pi  \sqrt{\Theta } M^2},
\end{split}
\fe
where $\lambda \equiv \sqrt{\left(54 \pi ^{3/2} a^2 M+54 \pi ^{3/2} M^3-432 \pi  \sqrt{\Theta } M^2\right)^2-108 \left(3 \pi  a^2+3 \pi  M^2-16 \sqrt{\pi } \sqrt{\Theta } M\right)^3}$.

Within the photon shell, spherical photon orbits typically oscillate in the \(\theta\)--direction between the polar angles \(\theta_{\mp}\) \cite{o100}. However, at the shell's boundaries, located at \(r = r_{p\mp}\), these orbits become confined to the equatorial plane (\(\theta = \pi/2\)) and take on a planar trajectory \cite{o100, o101}.

The shadow's shape is determined by the black hole's spin, additional parameters, and the angle \(\theta_0\) at which the observer views the black hole relative to its spin axis. The overall size of the shadow is set by the black hole's mass \(M\). As a result, for an observer positioned at \(r_0 \to \infty\) and observing from an inclination angle \(\theta_0\), the shadow would appear as a dark silhouette against a bright background, mapped by the following celestial coordinates \cite{afrim}
\ie
\left\{ \tilde{\alpha}, \tilde{\beta} \right\} = \left\{ - \xi_{c} \csc \theta_{0}, \pm \sqrt{\eta_{c} + a^{2} \cos^{2} \theta_{0} - \xi^{2}_{c} \cot^{2}\theta_{0}   }    \right\}.
\fe

\begin{figure}
    \centering
     \includegraphics[scale=0.5]{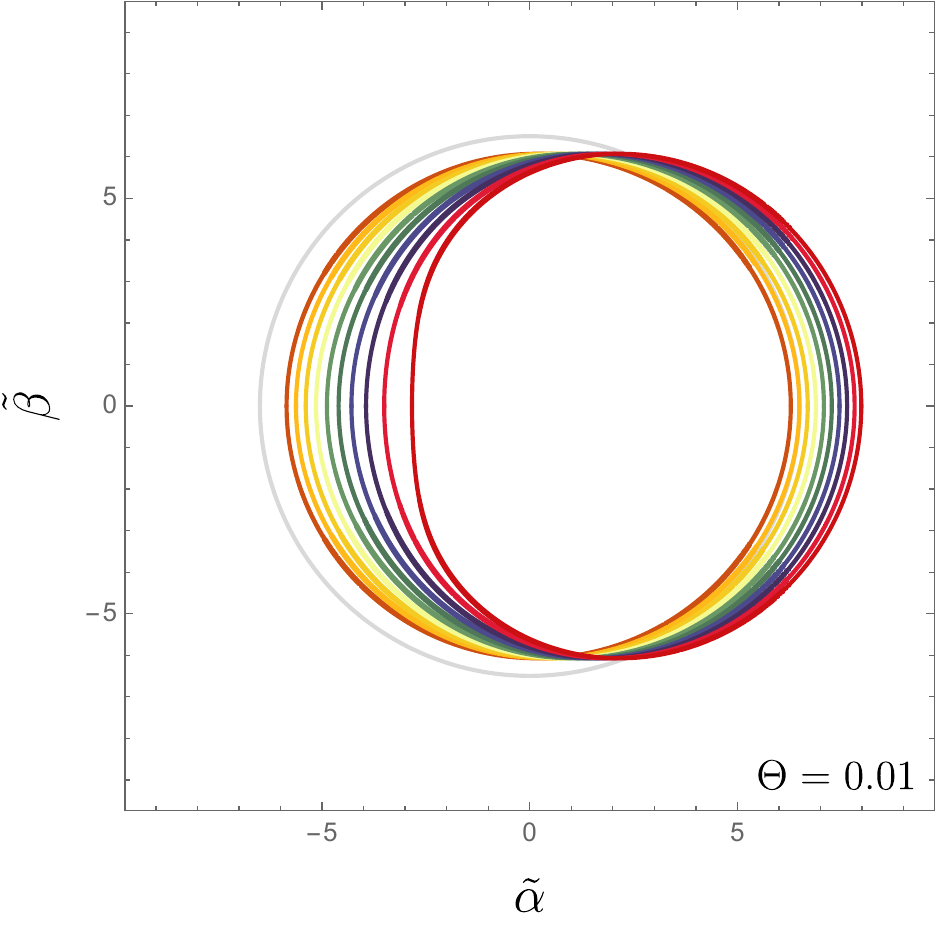}
      \includegraphics[scale=0.5]{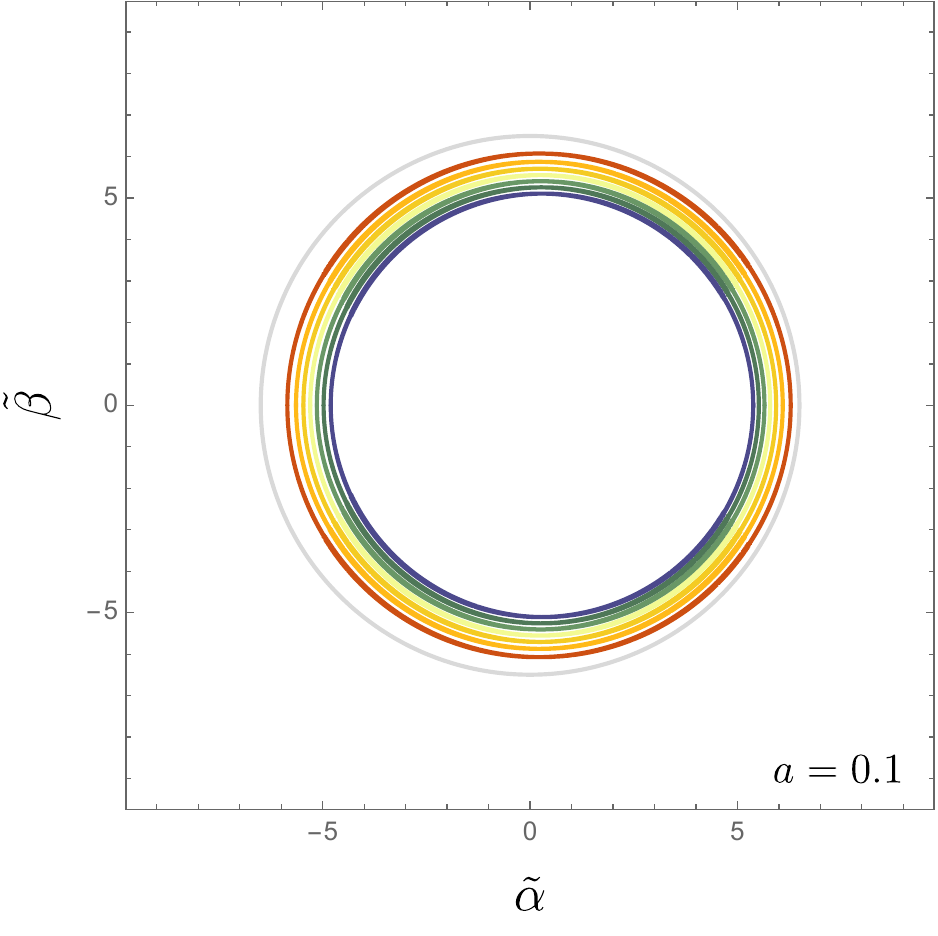}
    \caption{Shadows are display for different configuration of $\Theta$ and $a$.}
    \label{shadowsaa}
\end{figure}

To improve the clarity of our results, we present Fig. \ref{shadowsaa}, which illustrates the black hole shadows for a fixed value of the non--commutative parameter \(\Theta\) and various spin parameter configurations, \(a\) (on the left panel). Specifically, the gray line corresponds to the shadow of a Schwarzschild black hole, while the colored lines depict the progression of \(a\) values. From left to right, \(a\) increases from \(a=0.1\) to \(a=0.99\) in increments of 0.1, demonstrating how the shadow evolves with increasing the rotational parameter. On the other hand, we display the behavior of the shadows by varying the non-commutative parameter \(\Theta\) while keeping the spin parameter fixed at \(a=0.1\). Specifically, we consider \(\Theta = 0.01\), \(0.02\), \(0.03\), \(0.04\), \(0.05\), \(0.06\), and \(0.07\), with values increasing from left to right. Although the shadows for this configuration appear to be nearly perfect circles, they are in fact ellipses. This is particularly noticeable in the red ellipse, which deviates slightly from the gray circle, making it clearer that the colored ellipses are not true circles.

\section{Quasinormal modes \label{quasi}}

Quasinormal modes are essential for studying the stability and dynamic response of black holes to perturbations. In the context of non--commutative black hole solutions, their analysis becomes particularly relevant, as it reveals how the non--commutative parameter modifies the oscillatory frequencies and damping rates of these perturbations. This helps to understand the influence of noncommutative corrections on the spacetime geometry, connecting theoretical predictions with potential observational signatures, such as those detectable in gravitational wave signals. In this manner, this section is dedicated to investigating the \textit{quasinormal} modes of the metric (\ref{rotatingmetric}). To perform these calculations, we introduce a scalar perturbation and apply the method of separation of variables to derive the Teukolsky equations. The separation of angular and radial components allows us to first solve for the angular eigenvalue, \(A_{lm}\), from the angular equation. With the angular part resolved, we then proceed to the radial equation, which is properly formulated to enable the calculation of \textit{quasinormal} modes using the WKB approximation.


\subsection{The Teukolsky equations}

In his work, Teukolsky established that scalar, vector, and tensor perturbations in the Kerr spacetime all conform to a unified master equation for scalar variables of spin weight \(s\). Additionally, this master equation is solvable through separation of variables~\cite{teukolsky1972rotating}. We introduce \(\Psi\) as our scalar variable and represent the scalar wave as 
\begin{equation}
\label{eq6}
\Psi(t,r,\theta,\phi) = e^{-i \omega t }e^{i m \phi}\psi_r(r)\phi_{\theta}(\theta) \, ,
\end{equation}
so that the radial function $\psi_r(r)$ reads
\ie
\frac{\mathrm{d}^2 \psi_r}{\mathrm{d} r^2_*} + \left[ \frac{K^2-\Delta(r)\rho_{l m}}{(r^2+a^2)^2} \right] \psi_r=0 \,,
\quad
\frac{\mathrm{d}}{\mathrm{d}r_*} \equiv \frac{\Delta(r)}{r^2+a^2}\frac{\mathrm{d}}{\mathrm{d}r},
\fe
with
$
K = -\omega(r^2+a^2)+a m$, and $
\rho_{l m} =A_{lm}+a^2\omega^2-2am\omega$ (with $A_{lm}$ is the angular eigenvalue of this equation.). Additionally, let us consider \( A_{lm} = \left(l + \frac{1}{2}\right)^2 - \frac{a^2 \omega^2}{2} \left[1 - \frac{m^2}{\left(l + \frac{1}{2}\right)^2}\right] \) \cite{yang2012quasinormal}. Explicitly, the tortoise coordinate $r^{*}$ reads
\begin{eqnarray}
r^{*} &=& M \ln \left(\sqrt{\pi } \left(a^2-2 M r+r^2\right)+8 \sqrt{\Theta } M\right)-\nonumber\\&-&\frac{2 M \left(\sqrt{\pi } M-4 \sqrt{\Theta }\right) \tan ^{-1}\left(\frac{\sqrt[4]{\pi } (M-r)}{\sqrt{\sqrt{\pi } (a-M) (a+M)+8 \sqrt{\Theta } M}}\right)}{\sqrt[4]{\pi } \sqrt{\sqrt{\pi } (a-M) (a+M)+8 \sqrt{\Theta } M}}+r,
\end{eqnarray}
where if we consider $\Theta \to 0$ and $a \to 0$, we recover the tortoise coordinate of the Schwarzschild black hole, i.e., $ r^{*}_{s} = 2 M \ln (r-2 M)+r$.
In addition, we can define the effective potential as 
\ie
\begin{split}
& \tilde{\mathcal{V}} \equiv \frac{K^2-\Delta(r) \rho_{l m}}{(r^2+a^2)^2} \\
& =  \frac{\left(a m-\omega  \left(a^2+r^2\right)\right)^2-\left(\frac{1}{2} a \omega  \left(4 m \left(\frac{a m \omega }{(2 l+1)^2}-1\right)+a \omega \right)+l^2+l+\frac{1}{4}\right) \left(a^2+\frac{8 \sqrt{\Theta } M}{\sqrt{\pi }}-2 M r+r^2\right)}{\left(a^2+r^2\right)^2}.
\end{split}
\fe

An essential technique for identifying quasinormal modes, especially in black hole particle scattering scenarios, is the WKB method, originally introduced by Schutz and Will \cite{schutz1985black}. Over the years, this method has been significantly refined, with key improvements contributed by Konoplya \cite{konoplya2003quasinormal, konoplya2004quasinormal}. Its success relies on the potential having a barrier-like form, tapering to constant values as \( r^{*} \to \pm \infty \). By matching the series solution terms with the turning points at the potential peak, quasinormal modes are accurately determined. Under these conditions, Konoplya's formula is expressed as:
\ie
\frac{i(\omega^{2}_{n}-\tilde{\mathcal{V}}_{0})}{\sqrt{-2 \tilde{\mathcal{V}}^{''}_{0}}} - \sum^{6}_{j=2} \Lambda_{j} = n + \frac{1}{2}.
\fe
Konoplya's approach to calculating quasinormal modes involves several components. The term \( \tilde{\mathcal{V}}^{''}_{0} \) refers to the second derivative of the potential, evaluated at the peak \( r_{0} \), while the constants \( \Lambda_{j} \) are influenced by the effective potential and its derivatives at this maximum point. Notably, recent developments have pushed the WKB approximation to the 13th order, thanks to the work of Matyjasek and Opala \cite{matyjasek2017quasinormal}, which has led to a significant improvement in the precision of quasinormal frequency calculations.

An important aspect to highlight is that the quasinormal frequencies associated with the scalar field possess a negative imaginary part. This feature indicates an exponential decay over time, reflecting the process of energy loss through scalar wave emission. This behavior is consistent with previous studies on scalar, electromagnetic, and gravitational perturbations in spherically symmetric systems \cite{konoplya2011quasinormal,berti2009quasinormal,chen2023quasinormal}.

\begin{figure}
    \centering
    \includegraphics[scale=0.45]{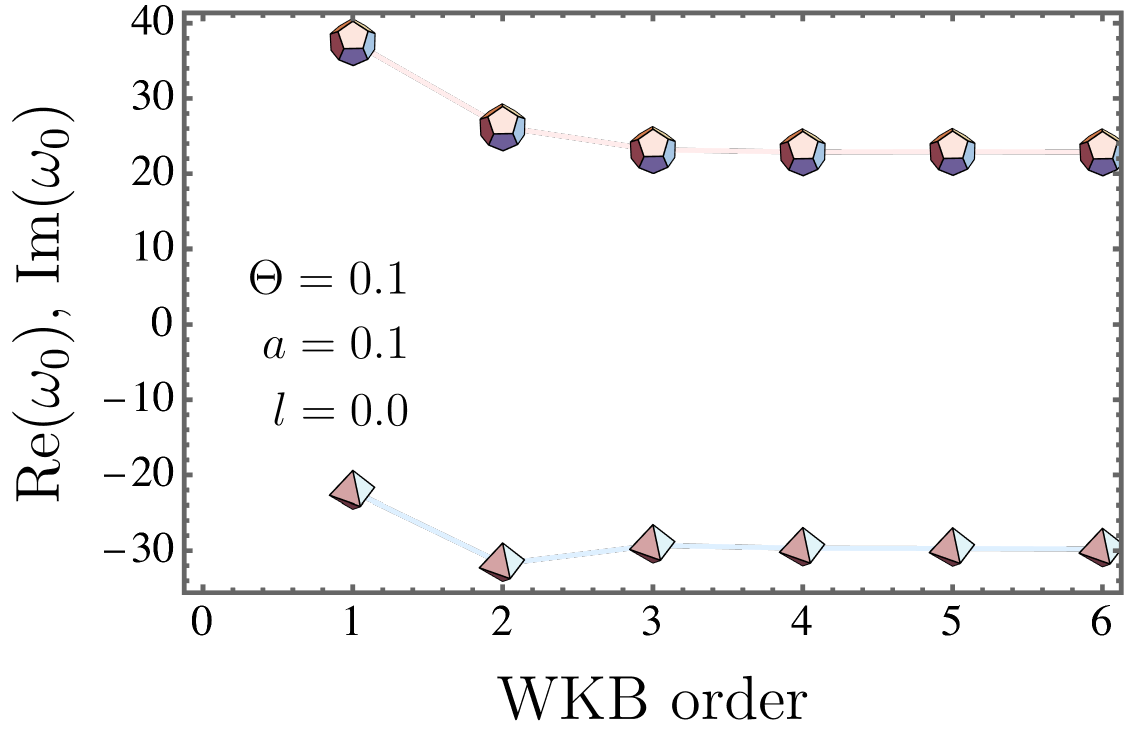}
    \caption{The convergence of higher--order terms in the WKB approximation}
    \label{wkbbehavior}
\end{figure}

Tables \ref{l0} and \ref{l1} display the behavior of the quasinormal modes as a function of \(\Theta\) and \(a\). For \(l = 0\), an increase in \(\Theta\) (with \(a\) held constant) results in modes that are overall less damped. A similar trend is observed when \(a\) increases. For the \(l = 1\) case, the system generally follows the same pattern, showing a comparable reduction in damping as in the \(l = 0\) scenario.
It is important to note that the WKB approximation exhibited good convergence at higher orders, as demonstrated in Fig. \ref{wkbbehavior}.

\begin{table}[!h]
\begin{center}
\caption{\label{l0}Utilizing the sixth--order WKB approximation, we demonstrate the quasinormal frequencies corresponding to varying values of \( a \) and \( \Theta \), specifically for \( l=0 \).}
\begin{tabular}{c| c | c | c} 
 \hline\hline\hline 
 \!\!\!\! $\Theta$ \,\,\,\,\,  $a$  & $\omega_{0}$ & $\omega_{1}$ & $\omega_{2}$  \\ [0.2ex] 
 \hline 
0.1 \, 1.0 & 157.511 - 3.965870$i$ & 157.252 - 11.9003$i$ & 156.733 - 19.8429$i$  \\ 
  0.2 \, 1.0 & 114.374 - 2.212940$i$  & 114.279 - 6.64014$i$  &  114.089 - 11.0713$i$  \\
 
 0.3 \, 1.0 & 90.6927 - 1.493640$i$ & 90.6418 - 4.48163$i$ & 90.5401 - 7.47179$i$ \\
 
0.4 \, 1.0 & 75.3692 - 1.103690$i$ & 75.3369 - 3.31151$i$ & 75.2724 - 5.52067$i$  \\
 
0.5 \, 1.0 & 64.5374 - 0.860429$i$ & 64.5148 - 2.58159$i$ & 64.4697 - 4.30366$i$ \\
 
  0.6 \, 1.0 & 56.4333 - 0.694995$i$ & 56.4165 - 2.08520$i$ & 56.3829 - 3.47606$i$  \\
 
0.7 \, 1.0 & 50.1230 - 0.575657$i$ & 50.1099 - 1.72713$i$ & 50.0837 - 2.87910$i$  \\
  
0.8 \, 1.0 & 45.0613 - 0.485793$i$ & 45.0507 - 1.45751$i$ & 45.0297 - 2.42960$i$ \\
 
   0.9 \, 1.0 & 40.9067 - 0.415872$i$ & 40.8981 - 1.24772$i$ & 40.8807 - 2.07987$i$  \\
  
 1.0 \, 1.0 & 37.4338 - 0.360047$i$ & 37.4265 - 1.08023$i$ & 37.4120 - 1.80065$i$ \\
   [0.2ex] 
 \hline \hline \hline 
  \!\!\!\! $\Theta$ \,\,\,\,\,  $a$  & $\omega_{0}$ & $\omega_{1}$ & $\omega_{2}$  \\ [0.2ex] 
 \hline 
 0.1 \, 0.1 & 29.3802 - 5.39054$i$ & 27.0592 - 16.7258$i$ & 22.8654 - 29.7885$i$ \\ 
 
 0.1 \, 0.2  & 57.5990 - 5.61321$i$ & 56.5543 - 16.9506$i$ & 54.4921 - 28.6304$i$ \\
 
 0.1 \, 0.3  & 82.6369 - 5.54126$i$ & 81.9381 - 16.6707$i$   &  80.5455 - 27.9434$i$  \\
 
 0.1 \, 0.4  & 103.580 - 5.39120$i$ & 103.039 - 16.1997$i$ &  101.960 - 27.0873$i$ \\
 
 0.1 \, 0.5  & 120.319 - 5.19552$i$ & 119.869 - 15.6031$i$ & 118.969 - 26.0606$i$ \\
 
 0.1 \, 0.6  & 133.206 - 4.96948$i$ & 132.815 - 14.9196$i$ & 132.034 - 24.9037$i$  \\
 
 0.1 \, 0.7  & 142.799 - 4.72482$i$ & 142.452 - 14.1823$i$ & 141.757 - 23.6635$i$ \\
 
 0.1 \, 0.8  & 149.696 - 4.47120$i$ & 149.383 - 13.4192$i$ & 148.757 - 22.3839$i$ \\
 
 0.1 \, 0.9  & 154.445 - 4.21631$i$ & 154.161 - 12.6528$i$ & 153.592 - 21.1012$i$  \\
  
 0.1 \, 1.0  & 157.511 - 3.96587$i$ & 157.252 - 11.9003$i$  & 156.733 - 19.8429$i$ \\
   [0.2ex] 
 \hline \hline \hline 
\end{tabular}
\end{center}
\end{table}

\begin{table}[!h]
\begin{center}
\caption{\label{l1}Utilizing the sixth--order WKB approximation, we demonstrate the quasinormal frequencies corresponding to varying values of \( a \) and \( \Theta \), specifically for \( l=1 \).}
\begin{tabular}{c| c | c | c} 
 \hline\hline\hline 
 \!\!\!\! $\Theta$ \,\,\,\,\,  $a$  & $\omega_{0}$ & $\omega_{1}$ & $\omega_{2}$  \\ [0.2ex] 
 \hline 
0.1 \, 1.0 & 51.0826 - 3.908890$i$ & 50.2380 - 11.7568$i$ & 48.5303 - 19.7003$i$  \\ 

  0.2 \, 1.0 & 37.5464 - 2.150120$i$  & 37.2248 - 6.46571$i$  &  36.5807 - 10.8284$i$  \\
 
 0.3 \, 1.0 & 30.2282 - 1.426660$i$ & 30.0475 - 4.28882$i$ & 29.6865 - 7.17790$i$ \\
 
0.4 \, 1.0 & 25.5873 - 1.033020$i$ & 25.4668 - 3.10477$i$  & 25.2260 - 5.19390$i$  \\
 
0.5 \, 1.0 & 22.3836 - 0.786788$i$ & 22.2952 - 2.36433$i$ & 22.1185 - 3.95389$i$ \\
 
  0.6 \, 1.0 & 20.0497 - 0.619194$i$ & 19.9810 - 1.86044$i$  & 19.8437 - 3.11034$i$   \\
 
0.7 \, 1.0 & 18.2846 - 0.498505$i$ & 18.2293 - 1.49761$i$ & 18.1186 - 2.50310$i$  \\
  
0.8 \, 1.0 & 16.9126 - 0.408053$i$ & 16.8669 - 1.22572$i$ & 16.7756 - 2.04813$i$ \\
 
   0.9 \, 1.0 & 15.8233 - 0.338237$i$ & 15.7850 - 1.01588$i$ & 15.7084 - 1.69706$i$  \\
  
 1.0 \, 1.0 & 14.9439 - 0.283129$i$ & 14.9115 - 0.850258$i$ & 14.8465 - 1.42003$i$ \\
   [0.2ex] 
 \hline \hline \hline 
  \!\!\!\! $\Theta$ \,\,\,\,\,  $a$  & $\omega_{0}$ & $\omega_{1}$ & $\omega_{2}$  \\ [0.2ex] 
 \hline 
 0.1 \, 0.1 & \text{Unstable} & 17.4318 - 131.031$i$ & 57.9126 - 627.369$i$ \\ 
 
 0.1 \, 0.2  & 15.5689 - 4.39546$i$ & 10.3331 - 15.9672$i$ & 5.07211 - 33.6574$i$ \\
 
 0.1 \, 0.3  & 24.0538 - 5.08446$i$ & 21.1477 - 16.0751$i$   &  16.2017 - 29.5696$i$  \\
 
 0.1 \, 0.4  & 31.3260 - 5.14566$i$ & 29.2878 - 15.8158$i$ &  25.4212 - 27.6952$i$ \\
 
 0.1 \, 0.5  & 37.2379 - 5.03672$i$ & 35.6295 - 15.3261$i$ & 32.4688 - 26.3091$i$ \\
 
 0.1 \, 0.6  & 41.8675 - 4.85452$i$ & 40.5166 - 14.7008$i$ & 37.8220 - 24.9877$i$  \\
 
 0.1 \, 0.7  & 45.3826 - 4.63498$i$ & 44.2077 - 13.9972$i$ & 41.8472 - 23.6549$i$ \\
 
 0.1 \, 0.8  & 47.9742 - 4.39697$i$ & 46.9320 - 13.2545$i$ & 44.8301 - 22.3151$i$ \\
 
 0.1 \, 0.9  & 49.8224 - 4.15233$i$ & 48.8874 - 12.5010$i$ & 46.9979 - 20.9898$i$  \\
  
 0.1 \, 1.0  & 51.0826 - 3.90889$i$ & 50.2380 - 11.7568$i$  & 48.5303 - 19.7003$i$ \\
   [0.2ex] 
 \hline \hline \hline 
\end{tabular}
\end{center}
\end{table}


\section{Conclusion\label{cccon}}

In this paper, we initiated our study by examining a spherically symmetric black hole within the framework of non--commutative geometry, considering the Lorentzian distribution of matter. By applying a modified Newman–Janis method, we derived a novel rotating black hole solution. We then explored the physical implications of this solution, particularly in terms of the event horizon structure, where both an inner \( r_{-} \) and an outer horizon \( r_{+} \) emerged. We found that increasing the non--commutative parameter \( \Theta \) and the rotation parameter \( a \) led to a noticeable attenuation in the horizon values.

The ergospheres were also studied, revealing an inner \( r_{e_{-}} \) and an outer configuration \( r_{e_{+}} \). Similar to the event horizons, the ergospheres displayed a shrinking behavior as \( \Theta \) and \( a \) increased. In addition, we examined the angular velocity, which showed a decrease in amplitude as \( \Theta \) increased, and a further reduction as \( a \) grew. The effects of non--commutativity on angular velocity were then compared to those of the standard Kerr solution.

For a thorough thermodynamic analysis, we calculated surface gravity, which was essential for determining thermal state quantities such as the \textit{Hawking} temperature, entropy, and heat capacity. Both the \textit{Hawking} temperature and entropy diminished as \( \Theta \) and \( a \) increased. Conversely, the heat capacity shifted toward higher values as these parameters grew. Also, all results were compared to the Kerr solution.

We also studied the geodesic motion, presenting numerical simulations in 3D for clarity. Special emphasis was placed on null geodesics and their corresponding radial accelerations. Furthermore, the photon sphere and the black hole shadows were examined in detail through analytical methods. Finally, we computed the quasinormal modes for scalar perturbations using the 6th--order WKB approximation. In general, we observed that increasing \( \Theta \) and \( a \) led to reduced damping in the quasinormal modes. A valuable direction for future study is to examine gravitational lensing in the black hole model derived here, following the approach in Ref. \cite{Vachher:2024ait}.




\section*{Acknowledgments}
\hspace{0.5cm}The authors would like to thank the Conselho Nacional de Desenvolvimento Cient\'{\i}fico e Tecnol\'{o}gico (CNPq) for financial support. P. J. Porf\'{\i}rio would like to acknowledge the Brazilian agency CNPQ, grant No. 307628/2022-1. The work by A. Yu. Petrov has been partially supported by the CNPq project No. 303777/2023-0. Moreover, A. A. Araújo Filho is supported by Conselho Nacional de Desenvolvimento Cient\'{\i}fico e Tecnol\'{o}gico (CNPq) and Fundação de Apoio à Pesquisa do Estado da Paraíba (FAPESQ), project No. 150891/2023-7. A. {\"O}. would like to acknowledge the contribution of the COST Action CA21106 - COSMIC WISPers in the Dark Universe: Theory, astrophysics and experiments (CosmicWISPers) and the COST Action CA22113 - Fundamental challenges in theoretical physics (THEORY-CHALLENGES). We also thank TUBITAK and SCOAP3 for their support. Furthermore, the authors express their gratitude to M. Ostroff for the valuable discussions and for providing the code used to compute the geodesic segment.

\bibliographystyle{ieeetr}
\bibliography{main}

\end{document}